\documentclass[fleqn,usenatbib]{mnras}

\usepackage[T1]{fontenc}

\DeclareRobustCommand{\VAN}[3]{#2}
\let\VANthebibliography\thebibliography
\def\thebibliography{\DeclareRobustCommand{\VAN}[3]{##3}\VANthebibliography}

\usepackage{graphicx}	
\usepackage{amsmath}	
\usepackage{amssymb}	
\usepackage{bm}
\usepackage[defaultcolor=red]{changes}

\usepackage{newtxtext,newtxmath}

\newcommand{\noopsort}[1]{}

\title[A Random Walk Model for Halo Triaxiality]{A Random Walk Model for Halo Triaxiality}

\author[Menker \& Benson]{
Paul Menker,$^{1,2}$\thanks{E-mail: pmenker@usc.edu}
and Andrew Benson,$^{2}$
\\
$^{1}$University of Southern California, 3551 Trousdale Pkwy, Los Angeles, CA 90089, USA\\
$^{2}$Observatories, Carnegie Institution for Science, 813 Santa Barbara Street, Pasadena, CA 91101, USA=
}

\date{Accepted XXX. Received YYY; in original form ZZZ}

\pubyear{2015}

\begin{document}
\label{firstpage}
\pagerange{\pageref{firstpage}--\pageref{lastpage}}
\maketitle

\begin{abstract}
We describe a semi-analytic model to predict the triaxial shapes of dark matter halos utilizing the sequences of random merging events captured in merger trees to follow the evolution of each halo's energy tensor. When coupled with a simple model for relaxation toward a spherical shape, we find that this model predicts distributions of halo axis length ratios which approximately agree with those measured from cosmological N-body simulations once constrained to match the median axis ratio at a single halo mass.  We demonstrate the predictive and explanatory power of this model by considering conditioned distributions of axis length ratios, and the mass-dependence of halo shapes, finding these to be in good agreement with N-body results. This model provides both insight into the physics driving the evolution of halo triaxial shapes, and rapid quantitative predictions for the statistics of triaxiality connected directly to the formation history of the halo.
\end{abstract}


\section{Introduction}\label{sec:introduction}

Dark matter halos, which form through the gravitational collapse of overdense regions of the universe, have been found to be significantly non-spherical via analysis of N-body simulations \citep{2002ApJ...574..538J,2005ApJ...618....1H,2006MNRAS.367.1781A,2007MNRAS.377...50H,2008MNRAS.391.1940M,2011MNRAS.416.1377V,2014MNRAS.443.3208D,2015MNRAS.449.3171B,2017MNRAS.466..181D,2017MNRAS.467.3226V}. Such halos have been found to be well-characterized as triaxial ellipsoids. Halos measured in N-body simulations are found to be to mostly prolate, although with a broad distribution of shapes, with lower mass halos being closer to spherical than higher mass ones \citep[e.g.][]{2002ApJ...574..538J,2015MNRAS.449.3171B}. The triaxial shapes of halos have also been found to exhibit weak trends with environment and halo formation history, with higher mass halos in low density environments (e.g. voids and walls of the cosmic web) being more prolate \citep[e.g.][]{2007MNRAS.378...55M,2017MNRAS.466.3834L,2021A&A...651A..56G,2021MNRAS.500.1029L,2021PhRvD.103f3517H}. These expectations from theoretical studies are broadly consistent with observational studies \citep[e.g.][]{2011ApJ...729...37M,2013SSRv..177..155L,2017MNRAS.467.4131V}.

This triaxial shape has important consequences for the orbital evolution of subhalos (and their associated satellite galaxies) leading to anisotropy in the spatial distribution of subhalos around their host \citep{2005ApJ...629..219Z,2007ApJ...671.1135K}, the strength of gravitational lensing exhibited by a halo \citep{2007MNRAS.380..149C}, cluster mass and concentration estimates \citep{2013SSRv..177..155L}, and two-point correlation functions \citep{2012MNRAS.424.2954V}. Understanding the origins and statistical distribution of halo shapes is therefore important to many astrophysical measurements.

In addition to these characterizations of halo triaxiality using measurements from N-body simulations, there has been some analytic work aimed at characterizing halo shapes, typically based upon the statistics of Gaussian random density fields \citep{1995ApJ...439..520E,2011MNRAS.416..248R}.

Previous work, utilizing N-body methods, exploring the effects of mergers on halo shapes \citep{2004MNRAS.354..522M,2019MNRAS.487..993D} has found the radial mergers tend to produce prolate halos while mergers with higher angular momentum tend to produce oblate halos, indicating that halo shape is directly affected by the dynamics of halo mergers.

In this work, we construct a model for predicting dark matter halo shapes using a different approach. We extend the semi-analytic framework of \cite{benson_random-walk_2020} and \cite{johnson_random_2021}, in which key properties (spin and concentration respectively) of halos are estimated by following the accumulation of conserved quantities (angular momentum and energy respectively) during each merger event in a halo's formation history. Specifically, in this work, we follow the accumulation of the energy tensor during each merger to predict the triaxial shape of each halo. Unlike angular momentum and the (scalar) energy, the energy tensor is not a conserved quantity. However, we will show that it is approximately conserved during mergers of halos to a sufficient level for this approach to succeed.

\cite{benson_random-walk_2020} and \cite{johnson_random_2021} demonstrated that this merger tree-based, semi-analytic framework can accurately predict the distribution of halo spins and the distribution and mass-dependence of halo concentrations, and provides a reasonable estimate of the auto-correlation over time of these quantities as halos evolve. This approach has the advantage (compared to other empirical and semi-analytic methods) that a halo's properties are predicted directly from its merger history. Since the statistical properties of such merger histories depend fundamentally on the power spectrum of cosmological density perturbations, and the cosmological model, this approach therefore connects the statistics of these key physical properties of halos to the underlying cosmological model, while also directly capturing correlations of these properties with the specific formation history of a given halo.

In this work we will explore some of these same issues when this framework is applied to the energy tensor to predict the triaxial shapes of halos. Our goal is to both test the hypothesis that the distribution of halo shapes can be directly related to halo merger histories, and to provide an approximate model for halo shapes which connects, through the merger history, to fundamental quantities such as the cosmological model and power spectrum. The remainder of this paper is organized as follows. In \S\ref{sec:methods} we describe how the existing framework is applied to the case of the energy tensor and halo triaxiality. In \S\ref{sec:results} we show the resulting distributions of halo triaxial shapes and compare to measurements from N-body simulations. In \S\ref{sec:discussion} we discuss the successes and limitations of this model, and in \S\ref{sec:conclusions} we present our conclusions.

\section{Methods}\label{sec:methods}

It is well-established \citep{1997ApJ...490..493N} that the density profiles of spherically-averaged cold dark matter halos can be accurately described by the Navarro-Frenk-White~(NFW) profile given by:
\begin{equation}
\rho(r)=\rho_* \left[ \frac{r}{R_\mathrm{s}}\left(1+\frac{r}{R_\mathrm{s}}\right)^2 \right]^{-1}.
\label{eq:NFW}
\end{equation} 
This distribution, which appears to hold at all scales \citep{2020Natur.585...39W}, describes a dense central cusp with $\rho \propto r^{-1}$, transitioning to a more diffuse outer region with $\rho \propto r^{-3}$ at a characteristic scale radius, $R_\mathrm{s}$. The parameter $\rho_*$ provides the overall normalization of the density profile. The scale radius is often defined via the relation $R_\mathrm{s} = R_\mathrm{vir}/c$ where $c$ is known as the concentration parameter, and $R_\mathrm{vir}$ is the virial radius, defined as the radius enclosing a certain mean density contrast\footnote{Throughout this work we define $\Delta_\mathrm{vir}$ using the spherical collapse model \protect\citep[see, for example, ][appendix A]{1996MNRAS.282..263E}.}, $\Delta_\mathrm{vir}$. Cold dark matter halos  are formed through a process of hierarchical merging \citep{1978MNRAS.183..341W} of smaller halos, the remnants of which can persist for cosmological timescales as `subhalos' orbiting inside the host halo into which they merged \citep{1999ApJ...524L..19M}. A useful representation of this hierarchical merging process is the ``merger tree''---a graph that tracks the history of these mergers, and which can be augmented by associating properties with each node (halo) in the tree, such as halo mass, virial radius, and orbital properties.

The majority of prior (semi-)analytic work assumes a spherical halo \cite[although see][]{2005ApJ...632..706L,2005MNRAS.360..203S}. It is well-established however that cold dark matter halos formed in cosmological simulations are significantly non-spherical, being better described as triaxial ellipsoids \citep{1988ApJ...327..507F,2007MNRAS.376..215B}. In this paper, we utilize and extend the merger tree-based approach of \cite{benson_random-walk_2020} and \cite{johnson_random_2021} to compute the triaxial shape of each halo in a merger tree by tracking the evolution of the energy tensor of each halo.  Specifically,  we update the energy tensor of each halo after each merging event in the tree, and find the resulting triaxial axis ratios.

In the remainder of this section we present N-body simulations of merging halos, and describe a simple model of how such mergers drive the evolution of triaxiality. In \S\ref{sec:simulations} we present a suite of idealized merger simulations of triaxial halos and examine the shapes of the resulting halos. Then in \S\ref{sec:Conserved} we define the energy tensor and how it is changed during halo mergers, and in \S\ref{sec:treeApplication} describe how these results are applied to entire merger trees. In \S\ref{sec:Corrections}, we describe the free parameters of our model--- one which is new, and two  which are identical to, and constrained by, the parameters of the \cite{johnson_random_2021} model. An important difference with respect to the prior works of \cite{benson_random-walk_2020} and \cite{johnson_random_2021} arises because the energy tensor, unlike the angular momentum \citep{benson_random-walk_2020} and the scalar energy \citep{johnson_random_2021}, is not a conserved quantity---we discuss this issue in \S\ref{sec:sphericalization}. In \S\ref{sec:comparison} we validate our model by comparing its predictions with the results of our N-body simulations. Finally, in \S\ref{sec:Trees}, we explain how these equations are utilized to follow the energy tensor, and triaxial halo shapes, through full merger trees.

\subsection{N-body Simulations}\label{sec:simulations}

To explore how mergers between halos drive the evolution of halo triaxiality we perform a suite of N-body simulations of isolated, merging halo pairs using the Gadget-2 code \citep{2005MNRAS.364.1105S}. Halos are modelled as triaxial exponentially-truncated NFW profiles \citep[][with $r_\mathrm{decay}=r_\mathrm{vir}$]{kazantzidis_2006} with isotropic velocity dispersions. We generalize the spherical NFW profile (eqn.~\ref{eq:NFW}) to a triaxial form, following, for example, \cite{2002ApJ...574..538J} we define a new coordinate, $m$, through the relation
\begin{equation}
  m^2(\mathbfit{x})=\sum_i\frac{x_i^2}{a_i^2},  
 \label{eq:coordinates}
\end{equation}
where the $x_\mathrm{i}$ are the position in a Cartesian coordinate system aligned with the principle axes of the triaxial halo, and $a_\mathrm{i}$ are the axis ratios of the triaxial halo, defined such that $a_1 > a_2 > a_3$, and we set $a_1 = 1$ for convenience with no loss of generality. The density at any point $\mathbfit{x}$ is then simply $\rho(m[\mathbfit{x}])$ with $\rho(r)$ being the spherical NFW profile as defined in equation~(\ref{eq:NFW}).

Initial conditions for these triaxial halos are generated using the Agama code \citep{2019MNRAS.482.1525V}. We modify Agama to support the specific truncated, triaxial NFW density profile described above, and approximate the corresponding potential using Agama's {\tt Multipole} potential class with maximum angular orders $l_\mathrm{max}=8$, $m_\mathrm{max}=6$. We then use Agama's Schwarzchild modeling capabilities to construct equilibrium particle distributions in a variety of triaxial potentials. For each halo, we generate a library of 200,000 orbits, each integrated for 50 orbital periods, and then solve for the weight of each orbit constrained to reproduce the target density profile and isotropic velocity dispersion. From this library we then sample 2,000,000 particle coordinates in the 6-d phase space, which are then sub-sampled as needed to construct our N-body halo realizations. For the subset of spherical halos that we consider we also generate initial conditions using Eddington's formula (as implemented in Galacticus; \citealt{binney_galactic_2008}, eqn.~4.46b). We find that the results of our simulations when using spherical halos are identical (to within statistical noise) for Agama or Galacticus-generated initial conditions).

\begin{table}
\caption{Parameters of idealized merger simulations. In this table, ``Label'' is used to identify each simulation in the text, ``$\xi$'' is the mass ratio of the merging halos, ``$\theta$'' is the angle between their initial relative velocity vector and the line connecting their centers, ``$c$'' is the concentration of the halo, and ``$a_2$'' and ``$a_3$'' are the initial axis ratios of the halo. These last three parameters, $(c,a_2,a_3)$, are listed separately for the primary and secondary halos.}
\label{tb:sims}
\begin{center}
\begin{tabular}{ccccccccc}
\hline
 & & & \multicolumn{3}{c}{\textbf{Primary}}& \multicolumn{3}{c}{\textbf{Secondary}} \\
\textbf{Label} & $\bm{\xi}$ & $\bm{\theta}$ & $\mathbf{c}$ & $\mathbf{a_2}$ & $\mathbf{a_3}$ & $\mathbf{c}$ & $\mathbf{a_2}$ & $\mathbf{a_3}$ \\
\hline
a & 0.1 & 0 & 10.0 & 1.0 & 1.0 & 10.0 & 1.0 & 1.0 \\
b & 0.1 & 0 &  9.0 & 1.0 & 1.0 & 12.0 & 1.0 & 1.0 \\
c & 0.1 & $\pi/8$ & 10.0 & 1.0 & 1.0 & 10.0 & 1.0 & 1.0 \\
d & 0.1 & $\pi/4$ & 10.0 & 1.0 & 1.0 & 10.0 & 1.0 & 1.0 \\
e & 0.3 & 0 & 10.0 & 1.0 & 1.0 & 10.0 & 1.0 & 1.0 \\
f & 1.0 & 0 & 10.0 & 1.0 & 1.0 & 10.0 & 1.0 & 1.0 \\
g & 0.1 & 0 & 10.0 & 0.7 & 0.7 & 10.0 & 0.5 & 0.5 \\
h & 0.1 & 0 & 10.0 & 1.0 & 0.7 & 10.0 & 1.0 & 0.4 \\
i & 0.1 & 0 & 10.0 & 0.8 & 0.7 & 10.0 & 0.5 & 0.4 \\
j & 0.3 & $\pi/4$ & 10.0 & 0.8 & 0.7 & 10.0 & 0.5 & 0.4 \\
k & 0.5 & $\pi/8$ & 10.0 & 0.5 & 0.5 & 10.0 & 0.5 & 0.4 \\
l & 0.2 & 0 & 10.0 & 1.0 & 1.0 & 10.0 & 0.5 & 0.5 \\
\hline
\end{tabular}
\end{center}
\end{table}

We consider numerous combinations of primary and secondary halos, as detailed in Table~\ref{tb:sims}, spanning a range of mass ratios, angles of approach, concentrations, and triaxialities.  The primary halo always has a mass of $10^{12}\mathrm{M}_\odot$. We use a particle mass of $10^7\mathrm{M}_\odot$ and a softening length of 500~pc. The two halos are initially separated by 300~kpc along the $x$-axis (approximately equal to the virial radius of the primary halo). Both primary and secondary halos are assigned randomly-selected orientations.  Each pair of merging halos is then allowed to evolve for 20~Gyr. We measure the energy tensor of the system (i.e. including all particles, from both halos) at each snapshot of the simulation, then find the eigenvectors and eigenvalues of the tensor allowing us to obtain the components of the tensor in the coordinate system corresponding to the principal axes of the system. We additionally determine axis ratios of the merger product by measuring the inertia tensor, including all particles within the approximate virial radius ($\approx R_\mathrm{v,1} (1 + \xi)^{1/3}$ where $R_\mathrm{v,1}$ is the pre-merger virial radius of the primary halo, and $\xi$ is the mass ratio of the merging halos) of the merger product, and then finding the eigenvalues of that inertia tensor. 

\begin{figure*}
\begin{tabular}{cc}
\includegraphics[width=85mm] {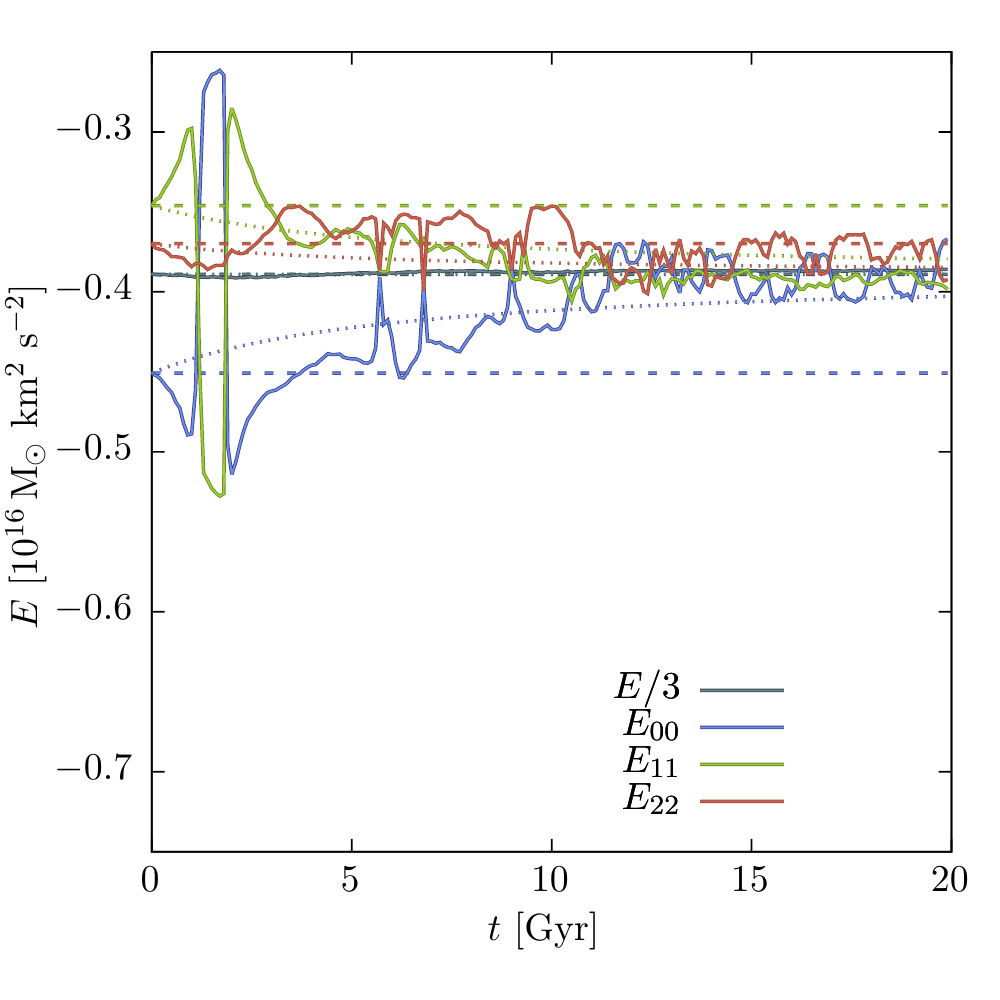} & \includegraphics[width=85mm] {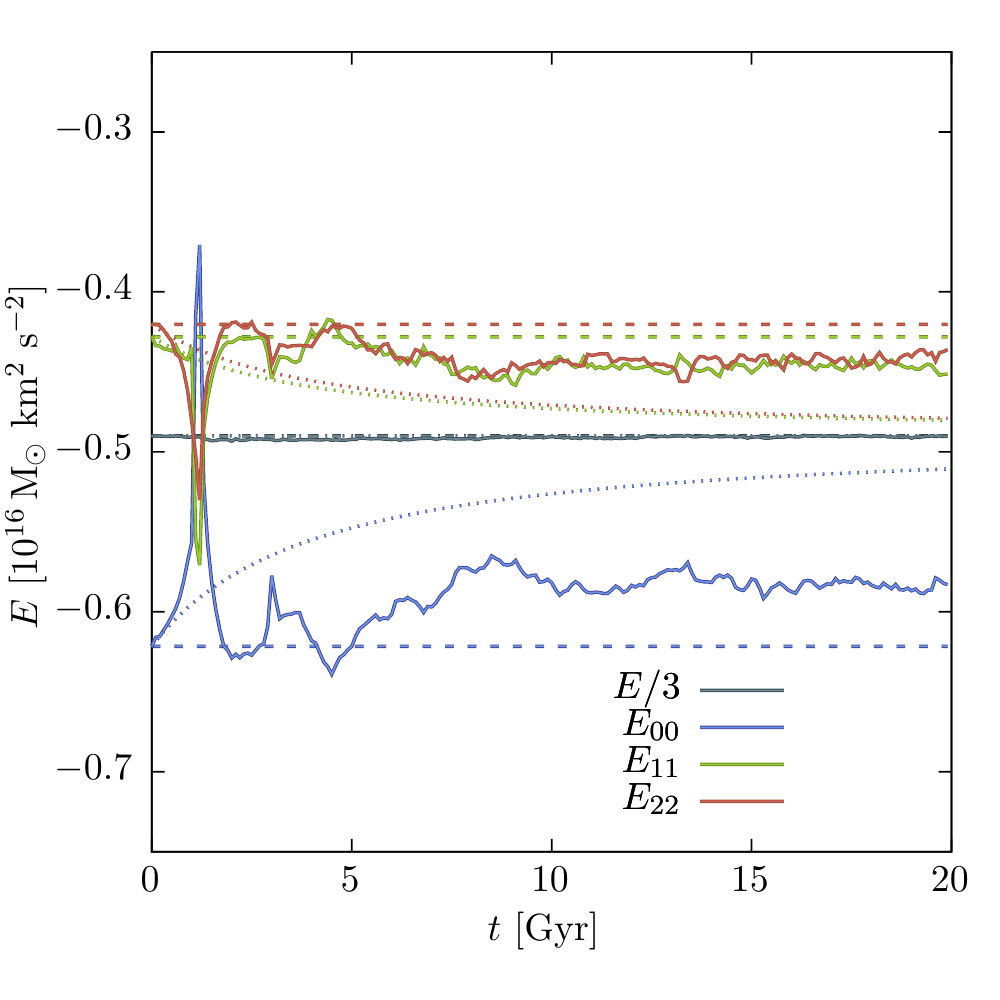}
\end{tabular}
\caption{Time evolution of the diagonal components of the energy tensor (and scalar energy shown as $E/3$ for comparison with the expectation for the diagonal components of the energy tensor for a spherical halo) in two N-body simulations of mergers between idealized dark matter halos is shown by the solid lines (numerical uncertainties on these lines---not shown for clarity---are $\Delta E \approx 0.01 \times 10^{16} \mathrm{M}_\odot \mathrm{km}^2\mathrm{s}^{-2}$ as determined from bootstrap resampling). Left and right panels show results for simulations d and g from Table~\protect\ref{tb:sims} respectively. The merging halos are initially separated along the $x$-axis and in the tangential merger case the initial velocity vector lies in the $x$--$y$ plane. Subscript indices 0,1,2 correspond to the $x,y,z$ axes respectively. Horizontal dashed lines indicate the initial value of each component of the tensor. Dotted lines show the expectations from our model, to be described in \S\protect\ref{sec:Conserved}.}
\label{fig:energyTensorSimulation}
\end{figure*}

Figure~\ref{fig:energyTensorSimulation} shows the time evolution of the diagonal components of the energy tensor in simulations d and g from Table~\protect\ref{tb:sims} in the left and right panels respectively. The scalar energy is also shown as $E/3$ for comparison with the expectation for the diagonal components of the energy tensor for a spherical halo. (Subscript indices 0,1,2 correspond to the $x,y,z$ axes respectively. Numerical uncertainties on these lines---not shown for clarity---are $\Delta E \approx 0.01 \times 10^{16} \mathrm{M}_\odot \mathrm{km}^2\mathrm{s}^{-2}$ as determined from bootstrap resampling.) Horizontal dashed lines indicate the initial value of each component of the tensor. Dotted lines show the expectations from our model, to be described in \S\ref{sec:Conserved}. We note that the total energy (shown by the grey lines) is conserved throughout the simulation. There are three distinct phases of evolution that can be identified in these results. Between 0 and 2~Gyrs, corresponding to the violent merging phase, the components of the energy tensor change rapidly. During this transitory phase the potential is changing rapidly, and the remnant, merged halo has not yet settled into a new equilibrium.

In the second phase, from around ~2 Gyrs to ~6 Gyrs, the remnant halo has settled into a new equilibrium, and the components of the energy tensor are relatively constant\footnote{In the left panel there are some fluctuations in the values of the energy tensor components---particularly noticeable around 2~Gyr where the $E_{00}$ and $E_{11}$ components switch values. This is a consequence of the eigenvalue decomposition that we perform. Due to the symmetry of the system and particle noise the labeling of axes (determined by the rank order of eigenvalues) can switch. This does not affect the physical attributes of the system, or our interpretation of the results.}. Importantly, they are constant at values close to their initial, pre-merger values. This is consistent with our assumption that the energy tensor is approximately conserved through the merger. 

Finally, from about 6~Gyr onward, the components of the energy tensor slowly converge towards $E/3$, indicating that the remnant halo is becoming more spherical. This process of ``\emph{sphericalization}'', in which the total energy remains fixed, but the components of the energy tensor equalize over timescales comparable to the age of the Universe is a consequence of the non-conserved nature of the energy tensor. Interestingly, sphericalization appears to happen on a timescale (around 5~Gyr) a few times longer longer than the dynamical time of the halo (which is of order 1~Gyr).  It is unclear if a large merger always resets the process of sphericalization, or if they can speed up the process in certain, coincidental circumstances.

The long-term (relative to the dynamical) stability of triaxial systems is well-studied. \cite{1991ARA&A..29..239D}, discussing equilibrium models for elliptical galaxies, note that ``the timescale for differential precession between neighboring orbits is usually of the order 5--10 dynamical times and becomes comparable to a Hubble time beyond a few optical radii.'' Similarly, \cite{2012MNRAS.419.3268V} perform simulations of the evolution of cuspy triaxial systems, finding that models with a weak ($\rho(r) \propto r^{-1}$) density cusp are remarkably stable, while steeper cusps ($\rho(r) \propto r^{-2}$) lead to evolution away from triaxiality due to chaotic diffusion of orbits. Similarly, \cite{1998ApJ...498..625M}, studying the effects of a central black hole on the shape evolution of triaxial galaxies, found that shape evolution occurs only on timescales of order hundreds of dynamical times in the limit of small black hole masses. These results are consistent with studies of cosmological N-body simulations which show that halos can maintain triaxial shapes over cosmological timescales \citep{1988ApJ...327..507F,1991ApJ...378..496D,2002ApJ...574..538J}.

\begin{figure*}
\includegraphics[width=190mm]{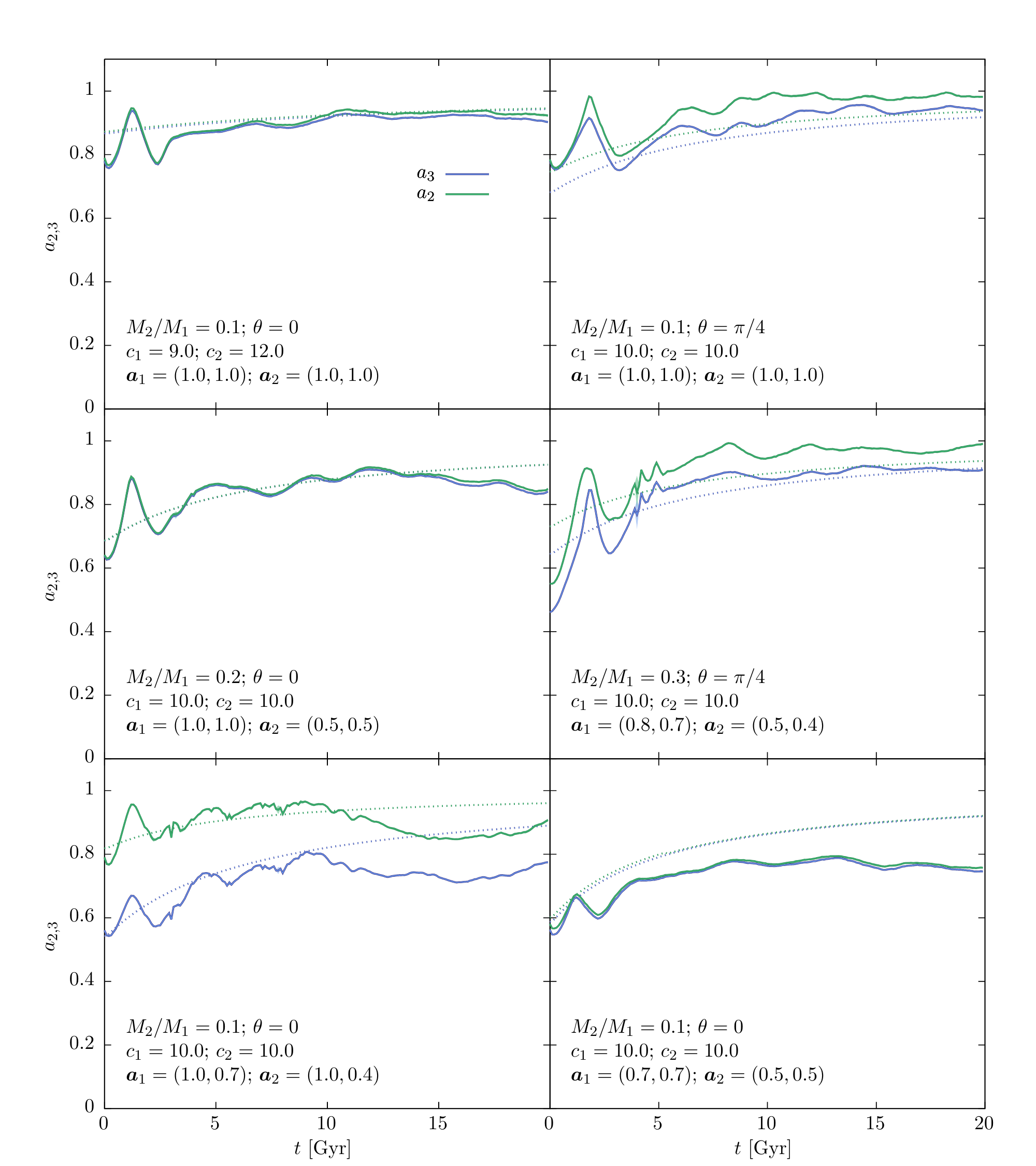} 
\caption{Axis ratios of the merged halo as a function of time in our idealized N-body simulations. Each panel shows results from a different simulation, with parameters as specified in the lower left corner of the panel. (Specifically, $M_2/M_1$ gives the mass ratio of the merging halos, $\theta$ the angle between the initial velocity vector of the halos and the line joining their centres, $c$ is the concentration of the primary or secondary halo as indicated by the subscript, and $\bm{a}$ is a vector consisting of $(a_2,a_3)$ of the primary or secondary halo as indicated by the subscript.) Green and blue solid lines show $a_2$ and $a_3$ measured from the simulation, while the corresponding dotted lines show results from our model (to be described in detail in \S\protect\ref{sec:comparison}).}
\label{fig:axisRatiosSimulation}
\end{figure*}

Figure~\ref{fig:axisRatiosSimulation} shows axis ratios, $a_2$ and $a_3$, measured from six of our simulations. Each panel shows the parameters of the simulation (mass ratio, angle of approach, concentrations and axis ratios of primary and secondary halos). Solid lines show the axis ratios measured directly from each simulation. Dotted lines indicate expectations from our model, to be discussed in \S\ref{sec:comparison}.

The axis ratios of the simulated merger remnants show the same transitory phase (0--2~Gyr) as was apparent in the energy tensors (Figure~\ref{fig:energyTensorSimulation}), during which time the axis ratios show large oscillations. After this initial phase, the axis ratios evolve more gradually, generally trending toward $a_2=a_3=1$ as the halo sphericalizes over time. An interesting exception is shown in the lower left panel, which the axis ratios actually decrease between 10 and 16~Gyr, before increasing again. The cause of this long-timescale behaviour in this simulation is unclear---examining the simulation snapshots suggests that fallback of ejected particles may contribute to this effect, although it is not clear why this does not affect other simulations.

\subsection{Energy Tensor Evolution}\label{sec:Conserved}

We define the energy tensor of a self-gravitating system (in this work, a dark matter halo) as
\begin{equation}
 \mathbfss{E} = \mathbfss{K} + \mathbfss{W},
\end{equation}
where $\mathbfss{K}$ is the kinetic energy tensor defined as \citep[][eqn.~4.240b]{binney_galactic_2008}:
\begin{equation}
    K_\mathrm{ij} = \frac{1}{2} \int \mathrm{d}^3x \rho(\mathbfit{x}) \overline{v_\mathrm{i} v_\mathrm{j}},
\end{equation}
and $\mathbfss{W}$ is the potential energy tensor defined as \citep[][2.19]{binney_galactic_2008}:
\begin{equation}
    W_\mathrm{ij} = - \int \mathrm{d}^3x \rho(\mathbfit{x}) x_\mathrm{i} \frac{\partial \Phi}{\partial x_\mathrm{k}}.
\end{equation}
Systems in equilibrium obey a tensor virial theorem, $2 \mathbfss{K} + \mathbfss{W} = 0$ \citep[][eqn. 4.248]{binney_galactic_2008}. The trace of the energy tensor is equal to the sum of the scalar potential and kinetic energies, and is therefore a conserved quantity (in a time-independent system). However, the energy tensor itself is not a conserved quantity. Nevertheless, we will show in \S\ref{sec:comparison} that it is approximately\footnote{The conservation is approximate during the merger due to interactions with the environment, including loss of mass from the system during the merger. A similar break down of precise conservation was also seen by \protect\cite{johnson_random_2021} in the scalar energy (see \S\protect\ref{sec:Corrections} for details). We will show in \S\protect\ref{sec:optimization} that the same corrections for this non-conservation that were found by \protect\cite{johnson_random_2021} also apply here.} conserved during mergers, although over cosmological timescales the energy tensor evolves to become more isotropic (see \S\ref{sec:sphericalization}). Therefore we proceed, for now, under the assumption that $\mathbfss{E}$ is constant for a given pair of merging halos and the resulting merged system.

Considering two dark matter halos, labelled by superscript (1) and (2), that are about to merge, we can write the total energy tensor of the system, $\mathbfss{E}^\mathrm{(tot)}$, as:
\begin{equation} 
\mathbfss{E}^\mathrm{(tot)}= \frac{\mathbfss{W}^\mathrm{(1)}}{2}+\frac{\mathbfss{W}^{(2)}}{2} + \mathbfss{V}^\mathrm{(12)} + \mathbfss{K}^\mathrm{(orb)},
\label{eq:totalEnergyRough}
\end{equation}
where we have used the tensor virial theorem to write $\mathbfss{E}^\mathrm{(i)} = \mathbfss{W}^\mathrm{(i)}/2$ for the energy tensor of each individual halo, and where  $\mathbfss{V}^\mathrm{(12)}$ represent the potential energy tensor due to the gravitational interaction of the two halos immediately prior to merging, and $\mathbfss{K}^\mathrm{(orb)}$ is the kinetic energy tensor associated with the orbital motion of the two halos immediately prior to merging.

Our choice to set $\mathbfss{E}^\mathrm{(i)} = \mathbfss{W}^\mathrm{(i)}/2$ differs from the approach taken in \cite{johnson_random_2021}. In that work the energy of each halo was found by summing potential and kinetic terms, with the kinetic energy found by solving the Jeans equation in spherical symmetry. As shown by \cite{1996MNRAS.281..716C}, this approach leads to virial ratios, $2K/|W|$, in good agreement with N-body simulations, even though cosmological N-body halos are found to deviate somewhat (10--20\%) from $2K/|W|=1$ (i.e. they are not in virial equilibrium). \cite{johnson_random_2021} further show that this approach results in good agreement with the total energies of cosmological halos.

Unfortunately, it is not possible in general to solve the Jeans equations for triaxial systems of the type considered in this work\footnote{\cite{2003MNRAS.342.1056V} find general solutions for the Jeans equation for triaxial St\"ackel models, but we are not aware of other solutions.}. As such, we resort to assuming that the (tensor) virial theorem holds, and thereby estimating the total energy tensor from the potential energy tensor, without the need to compute the kinetic energy tensor. Given the fact that cosmological halos are not precisely in virial equilibrium this will introduce some error, at the 10--20\% level, into our model. Given the other simplifying assumptions made we do not consider this to be a critical problem.

The potential energy tensor, $\mathbfss{W}$, can be found analytically for a triaxial NFW potential. 
 Defining \citep[][eq.~2.117]{binney_galactic_2008},
\begin{equation} 
\psi(m)= \int_0^m 2R_\mathrm{s}\frac{\rho^*}{(1+\frac{m'}{R_\mathrm{s}})^2}\mathrm{d}m^\prime 
\end{equation}
 where the triaxial coordinate $m^2$ is defined in equation~(\ref{eq:coordinates}),  we find
\begin{equation}
W^\prime_{ii} = -\pi^2 a_2^2a_3^2Ga_i^2A_i \int_0^{c R_\mathrm{s}}[\psi(m)-\psi(c R_\mathrm{s})]^2 \mathrm{d}m.
\label{eq:potentialEnergyTensor}
\end{equation}
This gives
\begin{equation}
\mathbfss{W}^{\prime \mathrm{(i)}}=-\gamma\frac{G\tilde{M}^2}{R_\mathrm{s}}\begin{pmatrix} A_1 & 0 & 0 \\ 0 & a_2^2A_2 & 0 \\ 0  & 0 &a_3^2A_3 \\ \end{pmatrix},
\label{eq:virial}
\end{equation}
\citep[][eqn.~93]{chandrasekhar_ellipsoidal_1987} where
\begin{equation}
\gamma=\frac{9}{4}\left[ 1-\frac{1}{(1+c)^2}-\frac{2 \log(1+c)}{1+c} \right],
\end{equation}
and
\begin{equation}
\tilde{M}=\frac{4\pi}{3}a_2a_3\rho^*R_\mathrm{s}^3,
\end{equation}
and the virial radius is $c R_\mathrm{s}$. $\mathbfss{W}^\prime$ indicates components of the tensor defined in the coordinate system corresponding to the principal axes of the triaxial halo, and $\tilde{M}$ can be thought of as a {\it scale} mass.

The $A_\mathrm{i}$ terms are defined as \citep[][\S II]{1962ApJ...136.1037C}:
\begin{eqnarray}
A_1 &=& \frac{2}{\sin^3\phi \sin^2\theta}[F(\theta,\phi)-E(\theta,\phi)], \nonumber \\
A_2 &=& \frac{2}{\sin^3\phi \sin^2\theta \cos^2\theta} \nonumber \\
    & & \,\,\, \left[E(\theta,\phi)-F(\theta,\phi)\cos^2\theta -\frac{a_3}{a_2}\sin^2\theta \sin\phi \right], \nonumber \\
A_3 &=& \frac{2}{\sin^3\phi \cos^2\theta} \left[ \frac{a_2}{a_3}\sin\phi-E(\theta,\phi) \right],
\end{eqnarray}
where $E()$ and $F()$ are the incomplete elliptic integrals
\begin{equation}
E(\theta,\phi)=\int_0^\phi (1-\sin^2\theta \sin^2\phi')^{1/2}\mathrm{d}\phi',
\end{equation}
and
\begin{equation}
F(\theta,\phi)=\int_0^\phi (1-\sin^2\theta  \sin^2\phi')^{-1/2}\mathrm{d}\phi',
\end{equation}
with
\begin{equation}
\sin \theta = \sqrt[]{\frac{1-a_2^2}{1-a_3^2}},\quad \mathrm{and}\quad \cos\phi = a_3.
\end{equation}

By construction, $\mathbfss{W}^{\prime \mathrm{(i)}}$ is diagonal in a coordinate system aligned with the principal axes of the triaxial halo. Generically, though, the two coordinate systems aligned with the principal axes of our two halos will not coincide with each other. We therefore chose to rotate the potential energy tensor $\mathbfss{W}'^{\mathrm{(2)}}$ into the coordinate system corresponding to the principal axes of halo (1). We assume no correlation between the principal axes of halos (1) and (2). Thus, $\mathbfss{W}^{\prime\mathrm{(1)}}$ and $\mathbfss{W}^{\prime\mathrm{(2)}}$ are both generated as in equation \ref{eq:virial}, and we then set $\mathbfss{W}^\mathrm{(1)}=\mathbfss{W}^{\prime\mathrm{(1)}}$ and $\mathbfss{W}^\mathrm{(2)}=\mathbfss{R} \mathbfss{W}^{\prime\mathrm{(2)}} \mathbfss{R}^\mathrm{T}$, where $\mathbfss{R}$ is a pitch-yaw-roll rotation matrix chosen such that the principal axes of halo (2) are isotropically distributed relative to those of halo (1).  

We now define a coordinate system with the first axis aligned with the vector connecting the centres of our two halos, and the remaining axes perpendicular to this with a rotation chosen uniformly at random. In this coordinate system the interaction tensor, $\mathbfss{V}^\mathrm{(12)}$, can be written as
\begin{equation}
V^\prime_\mathrm{ij} = -\frac{GM_1M_2}{c^\mathrm{(1)}R_\mathrm{s}^\mathrm{(1)}} \begin{pmatrix} 1 & 0 & 0 \\ 0 & 0 & 0 \\ 0  & 0 & 0 \\ \end{pmatrix},
\label{eq:potential}
\end{equation}
where $R_\mathrm{v}^\mathrm{(1)} = c^\mathrm{(1)}R_\mathrm{s}^\mathrm{(1)} $ is both the virial radius of halo 1, and the distance between halos at time of merging.
In this same coordinate system, the components of the kinetic energy tensor, $\mathbfss{K}^\mathrm{(orb)}$, are trivially related to the radial and tangential velocities of halo (2). We must then rotate the components of these tensors into the coordinate system aligned with the principal axes of halo (1). The components of the required rotation matrix, $\mathbfss{Q}$, can be found from the dot products of the basis vectors of our primed coordinate system, $\hat{\mathbfit{x}}^\prime$ and the coordinate system of halo (1), $\hat{\mathbfit{x}}^\mathrm{(1)}$,
\begin{equation}
Q_\mathrm{ij} \equiv \langle \hat{x}_\mathrm{i}|\hat{x}^\prime_\mathrm{j} \rangle ,
\end{equation}
which can be computed explicitly.

\subsection{Applications to Merger Trees}\label{sec:treeApplication}

In \S\ref{sec:Conserved}, we described the effect of single merger between two halos on the energy tensor of the merged halo. However, in the formation process of any current halo, a very large number of mergers will have taken place. To account for this, we create merger trees, which represent a ledger of every merging of two progenitor halos. The merger tree records the mass of each progenitor halo, and the time and ordering of every merging event between progenitor halos. We can then apply the model of \S\ref{sec:Conserved} to each merging event in order, recording the resulting values of $a_2$ and $a_3$ along with the full energy tensor for each halo. This energy tensor is then used in the next merger event involving the resulting merged halo. In this way, our model is applied recursively through the tree, until we reach the final, $z=0$ halo and compute its triaxial shape.

Fully-specifying this application of our model to a merger tree requires additionally specifying how to assign orbits to merging halos, how to assign energy tensors to halos in the tree that have no progenitors, and how to deal with subresolution accretion. These aspects are described in the remainder of this section.

In \S\ref{sec:beyondConservation} we will discuss additional physics beyond that described in \S\ref{sec:Conserved}, including possible non-conservation of the energy tensor through sphericalization and interactions with the environment.

\subsubsection{Merging Halo Orbits}

As in \cite{benson_random-walk_2020} and \cite{johnson_random_2021}, we assign orbital velocities to merging halo pairs by drawing values at random from the joint distribution of these velocities measured from cosmological N-body simulations. Specifically, we use the distribution functions measured by \cite{2015MNRAS.448.1674J}. As in \cite{benson_random-walk_2020} we introduce a weak correlation between the orbital angular momentum of merging secondary halos and the spin vector of the primary halo, $\bm{\lambda}_\mathrm{P}$, such that the angle between these two vectors is distributed as
\begin{equation}
P(\cos \theta) = \frac{1}{2} (1 + \eta | \lambda_\mathrm{P} | \cos \theta) ,
\end{equation}
where the parameter $\eta$ controls the strength of the correlation. \cite{benson_random-walk_2020} found a best fit value of $\eta = (6.6^{+11.3}_{-4.5}) \times 10^{-1}$ when constraining their model to match the distribution of halo spin parameters measured in N-body simulations. We use a value of $\eta = 0.16$ (consistent with the value from \citealt{benson_random-walk_2020} within the uncertainties) in this work based on an updated fit of the \cite{benson_random-walk_2020} model to data from the MultiDark Planck simulation suite \citep{2013AN....334..691R,2016MNRAS.457.4340K} which constrains $\eta < 0.65$ at 65\% confidence with a best-fit value of $\eta = 0.16$.

\subsubsection{Initial Conditions}\label{sec:initialConditions}

Due to the finite resolution of our merger trees there are always some halos which have no progenitors. Therefore, for these halos we can not determine their energy tensor based on an application of equation~(\ref{eq:totalEnergyRough}). Instead, we assume that these halos are spherically-symmetric, and set the scale radius of the NFW profile using the concentration model of \cite{2019ApJ...871..168D}. The merger trees used in this work have a mass resolution equal to $10^{-4}$ times the mass of the final, $z=0$ halo. We have checked that this resolution is sufficient that the statistics of the $z=0$ halo triaxial shapes are insensitive to our choice of initial conditions in these unresolved progenitor halos.

\subsubsection{Subresolution accretion}\label{sec:accretion}

The finite resolution of merger trees also means that not all growth in halo mass is due to merging events captured in the tree---there is some sub-resolution accretion which also leads to mass growth.

To account for this subresolution accretion we simply assume that the energy tensor of the halo grows in proportion to the characteristic virial energy of the primary halo, $E_\mathrm{vir} \propto - \mathrm{G} M^2_\mathrm{vir} / R_\mathrm{vir}$, where $M_\mathrm{vir}$ and $R_\mathrm{vir}$ are the virial radius and mass of the halo respectively. Given the resolution of our merger trees ($10^{-4}$ times the $z=0$ halo mass) we find that this subresolution accretion has only minimal effects on our results.

\subsection{Beyond Energy Tensor 
Conservation}\label{sec:beyondConservation}

So far, we have considered the idealized case in which the energy tensor is perfectly conserved during, and between merger events. In this subsection we explore processes which break this assumption, and extend our model to account for them. These additions do not change how our model is applied to a merger tree, as outlined in \S\ref{sec:treeApplication}.

\subsubsection{Corrections due to interactions with the environment}\label{sec:Corrections}

The model described in \S\ref{sec:Conserved} correctly describes the merger of two smooth, NFW, isolated halos if all components of the energy tensor are conserved, and their gravitational interaction can be well-approximated by the interaction of two point masses. However, as was discussed in \S\ref{sec:introduction}, this is not the case. Even in the prior work of \cite{benson_random-walk_2020} and \cite{johnson_random_2021}, which worked with quantities that are precisely conserved in isolated systems (angular momentum and energy), we found the need to introduce terms breaking this conservation. First, cosmological halos, are not perfectly smooth, nor perfectly described by a triaxial NFW profile, and nor is their interaction precisely that of two point masses. These are approximations made in our model, and we expect some order unity differences between our model predictions and direct N-body simulations as a result. Furthermore, in practice, pairs of merging halos are not isolated systems---they interact with their environment, exchanging angular momentum, energy, and mass. We briefly review the corrections introduced by \cite{johnson_random_2021} to account for this below.

These corrections will be included both when comparing our model to idealized halo merger simulations in \S\ref{sec:comparison}, and when calibrating our model to cosmological simulations in \S\ref{sec:optimization}. While the ``environment'' of a merging pair of halos differs in these two cases---being a vacuum for the case of the idealized mergers, and a cosmological distribution of matter and halos in the cosmological case---interactions with the environment happen nevertheless. Specifically, even for the idealized simulations particles can be scattered onto unbound (or close to unbound) orbits, and lost from the system to the environment, carrying away energy. For this reason, we apply the model described below for non-conservation effects in both cases, although we allow for the possibility that the quantitative details will differ by retuning the parameters of the model in each case.

In addition to these environmental exchanges, in the current work we must account for the fact that the energy tensor is not conserved even in an isolated, post-merger system, for both idealized and cosmological calculations. We describe how this is accounted for in \S\ref{sec:sphericalization}.

To incorporate the effects of interactions with the environment we employ the same model as \cite{johnson_random_2021}. Specifically, we modify equation~(\ref{eq:totalEnergyRough}) to become
\begin{equation}
\mathbfss{E}^\mathrm{(tot)}= \frac{\mathbfss{W}^\mathrm{(1)}}{2}+\left[ \frac{\mathbfss{W}^{(2)}}{2} + \mathbfss{V}^\mathrm{(12)} + \mathbfss{K}^\mathrm{(orb)} \right] \left(1+\frac{M_2}{M_1}\right)^{-\gamma} (1 + b). \label{eq:energyFinal}
\end{equation}
The new term multiplies the kinetic, potential, and interaction energy of halo (2) but does not affect halo (1).  This is because it is only the \emph{added} energy of the merger that we are modifying. This term contains two free parameters, $\gamma$ and $b$, both of which we set to their best-fit values from \cite{johnson_random_2021}: $\gamma = 1.518$ and $b = 0.673$.

The form of this correction is somewhat arbitrary. In \protect\cite{johnson_random_2021}, we hypothesized that the failure of (scalar) energy conservation would depend on the mass ratio of the merging halos as the mass fraction which becomes unbound and lost from the system (thereby directly leading to a failure of energy conservation in the final halo) is strongly mass-dependent. This motivates the form of the $(1+M_2/M_1)^\gamma$ term---in the limit of $M_2 \ll M_1$ it tends to $1$ (i.e. perfect conservation) while allowing for either an increase or decrease in the energy for cases where $M_2 \sim M_1$. The precise form was chosen as a simple parameterization which captures this desired behaviour. Any mass-independent failure of conservation is then accounted for by the $(1+b)$ term. Here we implicitly assume that this mass-independent failure of conservation is also independent of all other halo properties, and so this term is a constant. The justification for this choice is that it is the simplest possible assumption, and allowed for a successful match to N-body measurements of halo concentrations in \cite{johnson_random_2021}.

\subsubsection{Sphericalization}\label{sec:sphericalization}

Assuming that sphericalization of the shape of our halos is driven by differential precession, we can estimate the timescale for this process. \cite{2016MNRAS.461.1590E} derive that the precession frequency in a flattened potential is given by
\begin{equation}
 \omega = \Omega(1-1/q),
 \label{eq:precession}
\end{equation}
where
\begin{equation}
\Omega = \sqrt{\frac{\mathrm{G} M_\mathrm{vir}}{r_\mathrm{vir}^3}}
\end{equation}
is the characteristic orbital frequency at the virial radius, and we estimate the flattening parameter $q$ as
\begin{equation}
q = (a_2a_3)^{-1/4}.
\label{eq:precessionShape}
\end{equation}
(Note that \cite{2016MNRAS.461.1590E} define the flattening in terms of the potential, rather than in terms of the density profile as in this work. The above expression accounts for the fact that the potential is less flattened than the density) Between each merger event in our merger tree we allow the eigenvalues of the energy tensor to evolve according to:
\begin{equation}
\dot{\lambda}_\mathrm{i} = -\alpha \omega (\lambda_\mathrm{i} - \bar{\lambda}),
\label{eq:sphericalization}
\end{equation}
where $\bar{\lambda}$ is the mean of the eigenvalues, and $\alpha$ is a dimensionless parameter. This approach ensures that the total energy ($\sum \lambda_\mathrm{i} \equiv 3 \bar{\lambda}$) is unchanging while driving the system toward a spherical shape. 

We note that the form of equation~(\ref{eq:sphericalization} implies that halos will be driven toward a spherical shape over cosmological timescales, unless $\alpha \ll 1$. This may be in conflict with the results of prior work, such as that of \cite{2012MNRAS.419.3268V} discussed in \S\ref{sec:simulations}.

However, because of the halo shape dependence in the precession timescale (equations~\ref{eq:precession} and \ref{eq:precessionShape}) the timescale for sphericalization becomes longer as the halo becomes more spherical---as such, complete sphericalization is less rapid than might be expected from the dynamical timescale and value of $\alpha$ alone. Second, in the cosmological case of most interest here, mergers will continuously drive the halo away from spherical, such that what matters most is the application of equation~(\ref{eq:sphericalization}) during the short intervals between frequent merger events---the long term behaviour implied by equation (23) is less important to determining the results of our model.

Nevertheless, it should be kept in mind that the above model for sphericalization is clearly an oversimplification. Based on the idealized merger simulation results of \S\ref{sec:comparison} applying this model over times of 10~Gyr or more (with no ongoing merger activity to drive the halo shape evolution) is unlikely to be accurate.

\subsection{Comparison with N-body Simulations}\label{sec:comparison}

Having developed our model for the evolution of halo axis ratios after a halo--halo merger, we can examine the accuracy of this model by comparing to the idealized N-body simulations presented in \S\ref{sec:simulations}.

In Figure~\ref{fig:energyTensorSimulation}, which shows the evolution of the components of the energy tensor in two of these idealized simulations, the dotted lines show the predictions of our model. Specifically, we used the eigenvalues of the energy tensor measured in the initial conditions of each simulation as the starting point for our model calculation. We then apply the model described in equation~(\ref{eq:sphericalization}) to account for the effects of sphericalization as the simulation progresses. This gives the eigenvalues of the energy tensor at each timestep of the simulation. We then solve equation~(\ref{eq:virial}) at each timestep to find the corresponding axis ratios predicted by our model.

Our model has three free parameters, $(b, \gamma, \alpha)$. We choose the optimal values of these parameters to minimize the median absolute deviation between our model prediction and the simulation-measured axis ratios across all of our idealized simulations, excluding the first 5~Gyr of each simulation to remove the transitory phase where the eigenvalues of the energy tensor and axis ratios show large fluctuations. We find optimal values of $(b,\gamma,\alpha)=(0.295,1.175,2.145)$. We note that these values are optimal for these idealized simulations, but will not necessarily be the optimal choice for cosmological halos. Therefore, we will re-visit tuning of these parameters when we apply our model to cosmological cases. For the values of $(b,\gamma,\alpha)$ given above, the median absolute deviation between simulation and model axis ratios ranges between 0.010 and 0.125, with a mean of 0.042. Therefore, our model expectations are within $\Delta a \approx 0.04$ of the simulation axis ratios 50\% of the time.

We begin by considering the left panel of Figure~\ref{fig:energyTensorSimulation}, which shows results from simulation ``d'', a 10:1 mass ratio merger with angle of approach $\theta = \pi/4$ between spherical halos. In this case, our model accurately captures the evolution of the components of the energy tensor with time. In particular, the non-conservation of these components due to sphericalization is clearly seen in this figure. Recall that the horizontal dashed lines indicate the initial values of each component, and so show the expectation if the energy tensor were perfectly conserved, with no sphericalization. The N-body results (solid lines) clearly deviate from this expectation, and this deviation is accurately captured by our model. We find that the majority of our simulations show similarly good agreement with our model.

There are, however, some exceptions. In the right panel of Figure~\ref{fig:energyTensorSimulation} we show results from simulation ``g'', a radial 10:1 merger between two oblate halos. In these case some sphericalization can once again be seen. However, here the sphericalization occurs more slowly than predicted by our model, meaning that at late times our model will predict a more spherical halo than was found in the N-body simulation. This demonstrates that, while sphericalization seems to occur in all of our simulations, our simple model (\S\ref{sec:sphericalization}) does not fully capture the details of this process, and so is an avenue for further exploration.

To judge the validity of our model in matching the simulation results for the energy tensor evolution we compute median absolute deviations (MADs) between the diagonal components of the tensor as measured from the simulation and as predicted by our model, normalized to $|E|/3$. For the optimal values of $(b,\gamma,\alpha)$ given above we find that the MAD ranges from $0.03$ in the best case (simulation ``b'' in Table~\ref{tb:sims}), to $0.22$ in the worst case (simulation ``h''). Two thirds of the simulations (8 out of 12) have a MAD of $0.10$ or smaller. In the majority of cases, our model assumptions (an energy tensor that is conserved through the merger and which then undergoes slow sphericalization) result in a match to the simulation results to within 10\%.

We next examine our model predictions for the axis ratios directly. Figure~\ref{fig:axisRatiosSimulation} shows axis ratios measured from six of our idealized simulations as solid lines, with dotted lines indicating the expectation from our model. The panels of Figure~\ref{fig:axisRatiosSimulation} are arrange in order of increasing maximum absolute deviation between the simulation and model axis ratios (from left to right, top to bottom).

The upper left panel shows results for a 10:1 radial merger between two spherical halos of differing concentrations. The high degree of symmetry in this case implies that $a_2=a_3$ should be expected. This is correctly predicted by our model, and is approximately true in the simulation also. After the initial transitory phase our model accurately matches the results of the simulation, including the gradual sphericalization. The middle left panel shows a similar result, in this case for a 5:1 mass ratio and an oblate secondary halo. Again our model accurate matches the N-body axis ratios and the slow sphericalization up to around 15~Gyr. After this time the simulated halo begins to become less spherical, causing a deviation from our model expectation. As discussed in \S\ref{sec:simulations} the cause of this late-time behaviour remains unclear, although examination of the simulation snapshots suggests that re-accretion of ejected particles may play a role.

The top and middle right panels show cases with a lower degree of symmetry, due to an angle of approach of $\theta=\pi/4$. The top right panel shows a 10:1 mass ratio merger between spherical halos, while the middle right panel shows a 3:1 mass ratio merger between two triaxial halos. In these cases our model is slightly offset, overestimating $a_2$ and $a_3$, but nevertheless captures the rate of sphericalization, and the approximate difference between the axis lengths, $a_2-a_3$.

Finally, in the bottom row of Figure~\ref{fig:axisRatiosSimulation} we show two cases where our model fails more seriously. In the left panel we examine a 10:1 radial merger between two prolate halos. Our model performs very well in matching the N-body results up until a time of around 10~Gyr. After this time the simulation-measured axis-ratios decreases for around 6~Gyr before beginning to climb again. Such behaviour is not expected in our model. Once again, the cause of this behaviour is not clear to us, but may be related to re-accretion of ejected particles. In the bottom right panel we show a 10:1 radial merger between two oblate halos. In this case sphericalization seems to cease after around 7~Gyr, resulting in our model becoming a progressively worse match to the simulation results.

As a further test of our sphericalization model we set $\alpha=0$ (thereby resulting in no sphericalization), recompute the predictions from our model, and compare these new, non-sphericalized predictions to the results of the N-body simulations. We find that the median absolute deviations between simulated and model axis ratios increase from the range 0.010--0.125 for $\alpha=2.145$ to the range 0.048--0.463 for $\alpha=0$. Thus, ignoring sphericalization results in discrepancies between our model and the simulations increasing by a factor of around 4---indicating that sphericalization is an important component of our model.

In summary, our model captures the axis ratios and their evolution during sphericalization measured in N-body simulations in a majority of cases with reasonable accuracy (typical median absolute deviations of $\Delta a \approx 0.04$). The most common failure occurs at late times (10~Gyr or more after the merger) where some simulated halos show a a trend of becoming less spherical---a behaviour not captured in our model. In a minority of simulations the sphericalization process appears to be significantly slower than assumed in our model, resulting in the degree of triaxiality of the halo being underestimated by our model. Overall, while imperfect, this level of agreement is sufficient to allow us to make predictions for the statistical properties of axis ratios in ensembles of mergers with at least moderate accuracy.

\subsection{Calibration and Validation\label{sec:Trees}}

To calibrate and validate our model we compare to the results reported by \cite{2015MNRAS.449.3171B} from the Millennium XXL simulation, and by \cite{2012JCAP...05..030S} from the Millennium I \& II simulations. We generate a distributions of triaxial shapes by constructing 10,000 merger trees for halos with $z=0$ masses of $10^{10}$, $10^{11}$, $10^{13}$, $10^{13}$, $10^{14}$, $10^{15} \mathrm{M}_\odot$ utilizing the approach of \cite{2008MNRAS.383..557P} with best-fit parameters given by \cite{2019MNRAS.485.5010B}. In constructing these trees we match cosmological and power spectrum parameters to the Millennium XXL simulation \citep{2012MNRAS.426.2046A}. 

Each tree provides virial masses and radii for all halos, along with the cosmic time at which each halo exists, orbital velocities for each merger, and progenitor-descendent relationships between halos. For each tree, we initialize scale radii and shapes of all halos with no progenitors (i.e. those just above the resolution limit of $10^{-4} M_0$, where $M_0$ is the mass of the $z=0$ halo) as described in \S\ref{sec:initialConditions}. The energy tensor of each of these initial halos is then computed.

Working forward in time through the tree, at each merger event in the tree we compute the energy tensor of the merged halo using equation~(\ref{eq:energyFinal}), accounting for any accreted mass as described in \S\ref{sec:accretion}. The axis ratios, $a_\mathrm{i}$, are then determined from the eigenvalues of the energy tensor\footnote{This must be done numerically as the potential energy tensor depends on the $a_\mathrm{i}$ in a non-trivial way. In practice we tabulate the relation between the eigenvalues of the potential energy tensor and the axis ratios to allow rapid solution.}. Between each merger event in the tree we apply our sphericalization procedure as defined in equation~(\ref{eq:sphericalization}). At the end of this process we have determined the energy tensor, and axis ratios for every halo in the tree. For this work we examine only the properties of the final, $z=0$ halo in the tree, which we have verified are converged with respect to the tree mass resolution.

\cite{2015MNRAS.449.3171B} and \cite{2012JCAP...05..030S} both excluded unrelaxed halos with a large offset (5\% and 7\% of the virial radius respectively) between the position of the most bound particle and the centre of mass. We cannot apply precisely this selection to our halos, but instead, when comparing results with \cite{2015MNRAS.449.3171B} and \cite{2012JCAP...05..030S}, we exclude any tree which experienced a major (mass ratio of 4:1 or greater) merger within the last 1~Gyr of evolution as these late merging halos may not have fully virialized by $z=0$. Other cutoff times between 0.5 and 2~Gyrs were tested, but the results were similar in all cases.

\section{Results}\label{sec:results}

We begin, in \S\ref{sec:optimization}, by constraining the value of our sphericalization parameter, $\alpha$, by matching to the distribution of $a_3$ found in N-body simulations. With the calibrated model we then examine the resulting behaviour for a single tree and compare the distribution of halo shapes with those found in N-body simulations.

\subsection{Calibration}\label{sec:optimization}

We find that, for $\alpha \gtrapprox 10$, all $a_i \approx 1$, that is halos are driven to be almost perfectly spherical, confirming our expectation that $\alpha \sim 1$ will be required. To constrain the value of $\alpha$ we tune it to find the best match to the distribution of $a_3$ measured from cosmological N-body simulations by \cite{2015MNRAS.449.3171B}. Specifically, we applied our model to a sample of 10,000 merger trees, each with $z=0$ halo mass of $10^{14}\mathrm{M}_\odot$, using values of $\alpha$ ranging from 1.5 to 2.5 chosen based on a preliminary examination of the resulting $a_3$ distribution. For each value of $\alpha$ we computed the median $a_3$ across our 10,000 trees and compared computed a goodness of fit measure $\chi^2 = [\bar{a}_3(\alpha) - \bar{a}_3^*]^2$ where $\bar{a}_3(\alpha)$ is the median value of $a_3$ computed from our model for a given $\alpha$, and $\bar{a}_3^*=0.53$ is the median $a_3$ measured by \cite{2015MNRAS.449.3171B} for halos of this mass. We then seek the value of $\alpha$ that minimizes $\chi^2$. We find an optimal value of $\alpha = 1.75$.

In the remainder of this paper we therefore use $(b,\gamma,\alpha)=(0.673,1.518,1.75)$.

\subsection{Individual Tree Behavior}

\begin{figure}
\includegraphics[width=85mm]{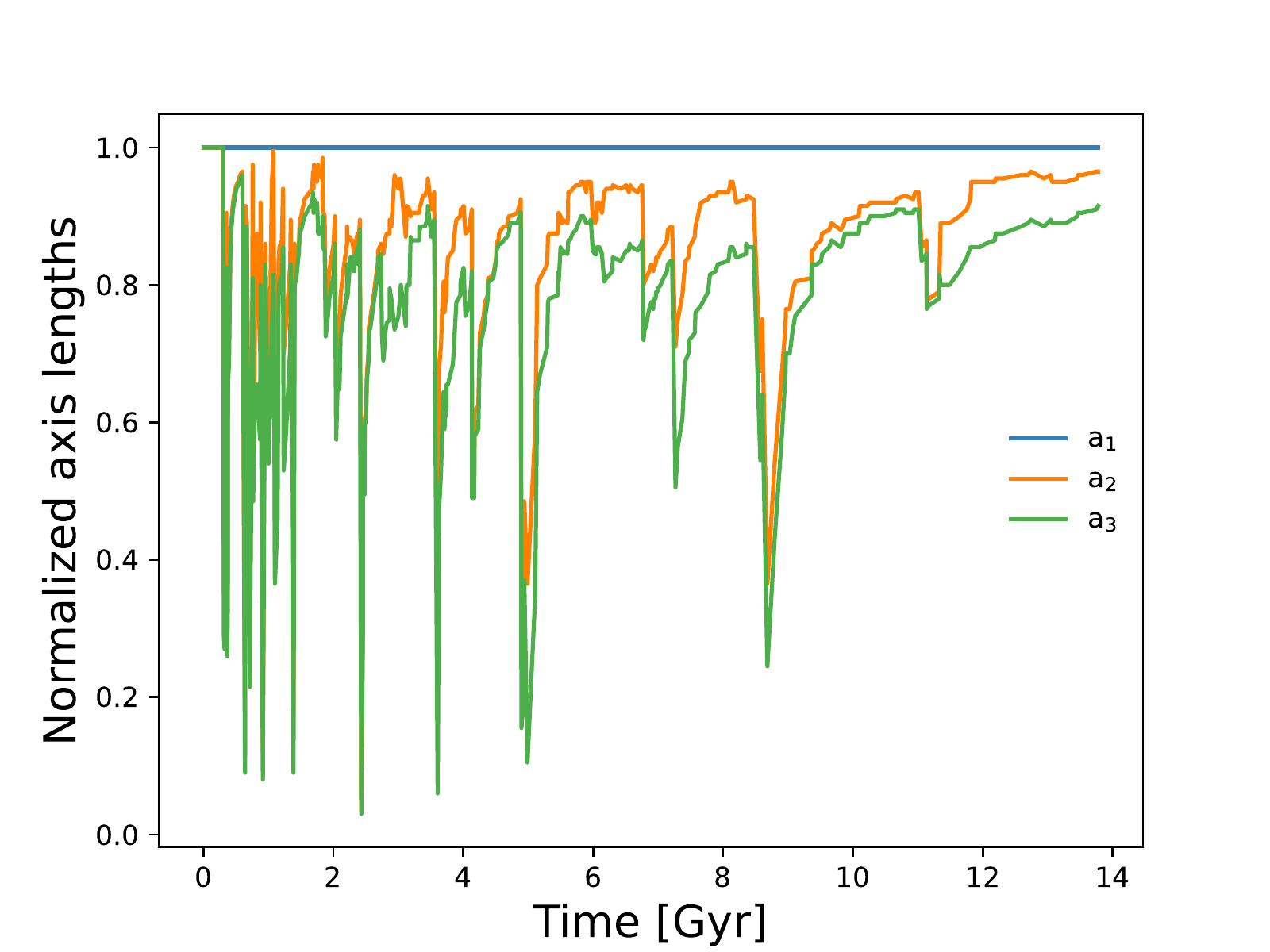}
\caption{The evolution with time of the axis lengths, $a_i$, in the main branch of a single, representative merger tree of mass $10^{14}\mathrm{M}_\odot$ at $z=0$ as described by our model. Note that $a_1=1$ by definition. Major merger events are clearly seen as sharp declines in $a_2$ and $a_3$, which then slowly drift toward 1 as a result of sphericalization and multiple, smaller mergers.}
\label{fig:singleHistory}
\end{figure}

Figure \ref{fig:singleHistory} shows the evolution of the axial ratios along the main branch of a single, representative tree of mass $10^{14}\mathrm{M}_\odot$ at $z=0$. By definition, $a_1$ is always 1---because the coordinate system is rotated after every merger to align with the principle axes of the halo, $a_1$ is always along the largest axis and should not be thought of as a particular direction in space. The other axis lengths, $a_2$ and $a_3$, show large variations---these are strongly correlated both because an individual merger event will affect both axes, and because they are both normalized to the principle axis length. After each major merger, the halo becomes very prolate, but sphericalization and the collective effects of many smaller mergers drive the halo back toward a more spherical configuration. Major mergers become more rare as the halo grows in mass and the timescale for growth (and, therefore, for changes in triaxial shape) increases as the halo evolves due to the decreasing density (and dynamical timescale) of the halos.

\subsection{Unconditioned Distributions of Axis Ratios}
 
\begin{figure*}
\begin{tabular}{cc}
\includegraphics[width=85mm]{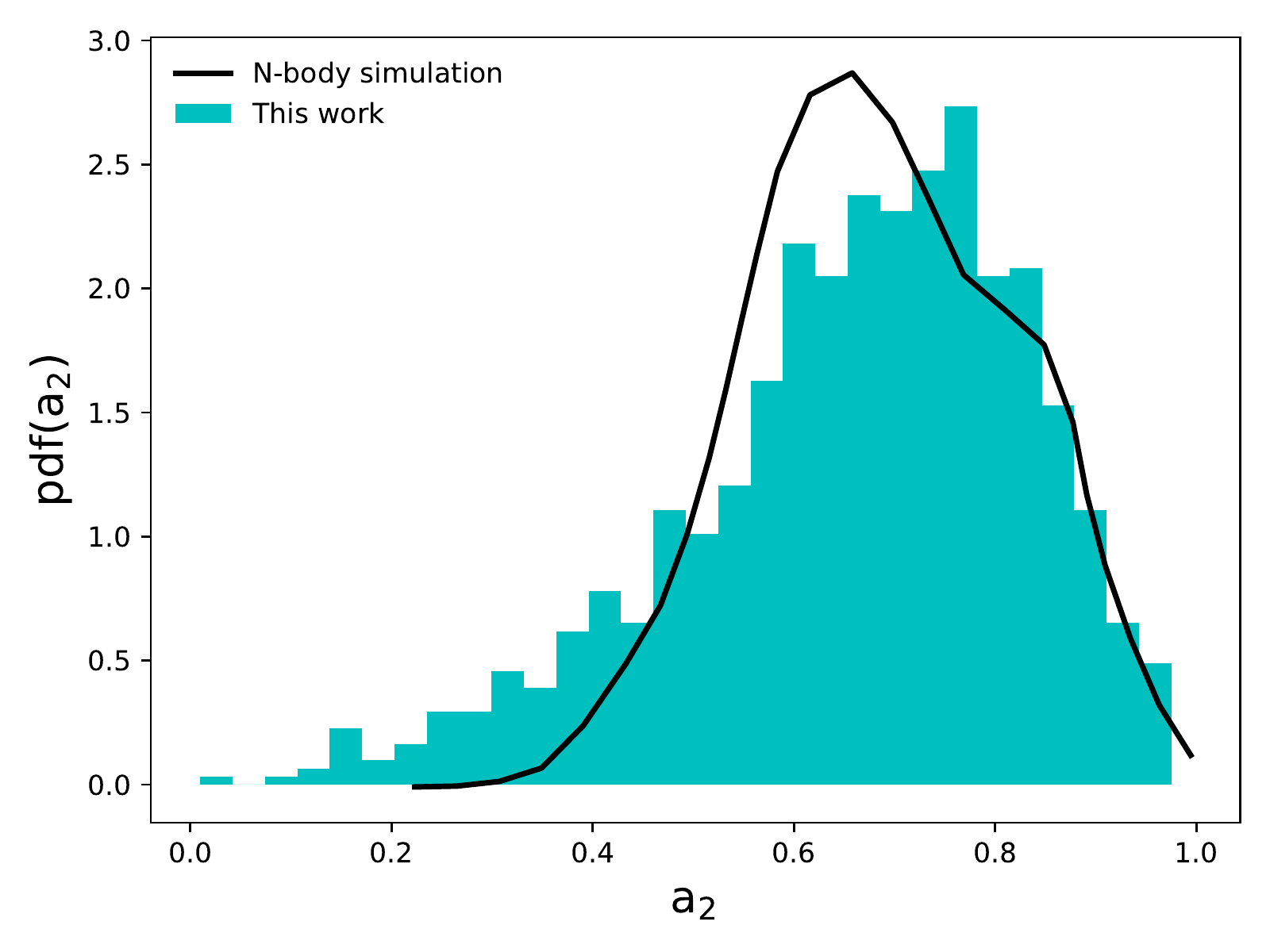} & \includegraphics[width=85mm]{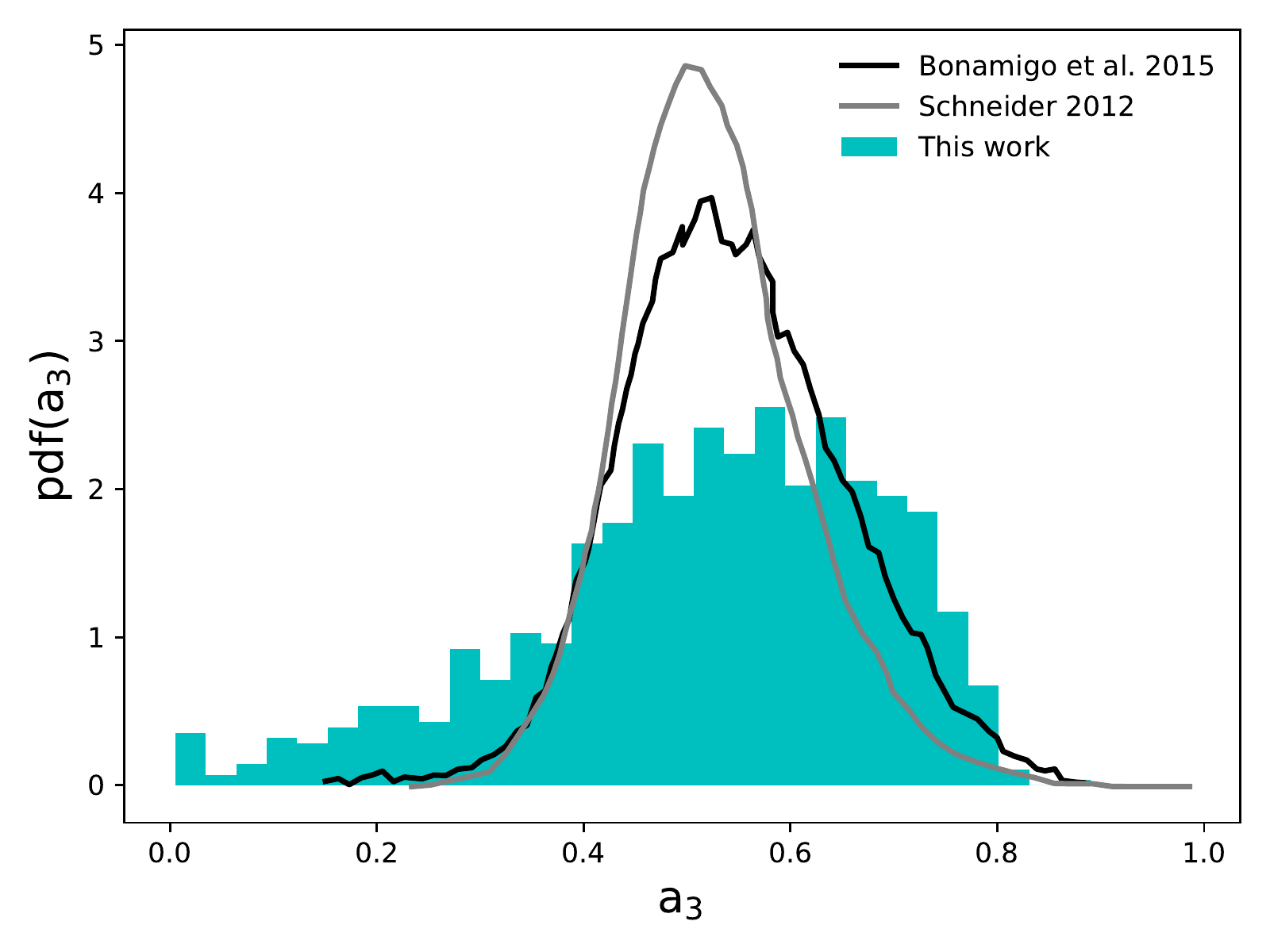}
\end{tabular}
\caption{Probability distribution functions for the axis ratios $a_2$ (left panel) and $a_3$ (right panel) derived by applying our model to 10,000 realizations of merger trees with mass $10^{14}\mathrm{M}_\odot$ at $z=0$.  In the left panel, for $a_2$, we overlay the corresponding result measured from cosmological N-body simulations by \protect\cite{2012JCAP...05..030S}. In the right panel, for $a_3$, we overlay N-body results from \protect\cite{2015MNRAS.449.3171B} and \protect\cite{2012JCAP...05..030S}}.
\label{fig:pdfUnconditioned}
\end{figure*}

\begin{figure}
\includegraphics[width=85mm]{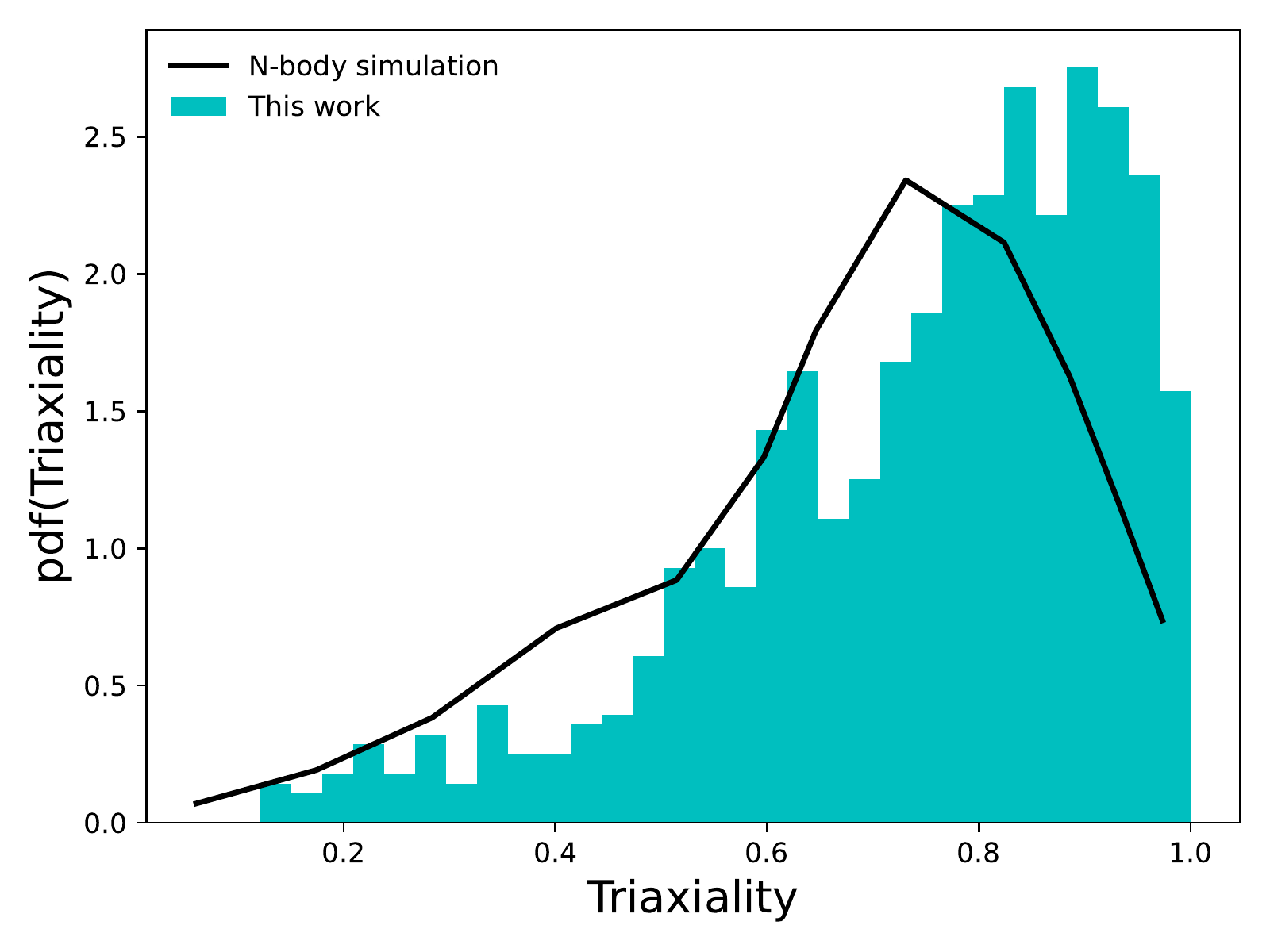}
\caption{Probability distribution function for the triaxiality parameter ($T \equiv [1-a_2^2]/[1-a_3^2]$) derived by applying our model to 10,000 realizations of merger trees with mass $10^{14}\mathrm{M}_\odot$ at $z=0$. N-body simulation results are taken from \protect\cite{2012JCAP...05..030S}.}
\label{fig:triaxiality}
\end{figure}

Figure \ref{fig:pdfUnconditioned} shows the distributions of $a_2$ and $a_3$ values (left and right panels respectively) derived by applying our model to 10,000 realizations of merger trees with mass $10^{14}\mathrm{M}_\odot$ at $z=0$. For the $a_2$ distribution, we overlay the corresponding result from \cite{2012JCAP...05..030S}\footnote{\protect\cite{2012JCAP...05..030S} uses a halo mass definition based on a density threshold of 200 times critical density, instead of the spherical collapse-motivated threshold used in this work and by \protect\cite{2015MNRAS.449.3171B}. In making the comparison in Figure~\ref{fig:pdfUnconditioned} we convert from this 200 times critical mass definition to a spherical collapse-motivated definition assuming NFW density profiles and the concentration model of \protect\cite{2019ApJ...871..168D}.}.  For the $a_3$ distribution, we overlay N-body results from both \cite{2012JCAP...05..030S} and \cite{2015MNRAS.449.3171B}. The agreement in the median value of $a_3$ with this N-body measurement is a result of our use of this statistic to tune the value of the sphericalization parameter, $\alpha$. Our predicted distributions for $a_2$ and $a_3$ show clear peaks with extended tails to low values. Comparing the distribution of $a_3$ with that from \cite{2015MNRAS.449.3171B} we find that our model predicts a broader distribution of $a_3$---in particular is has many more halos with very small values of $a_3 \lessapprox 0.3$. Our distribution of $a_3$ has 25 and 75 percentiles of $0.41$ and $0.64$, while that measured by \cite{2015MNRAS.449.3171B} has $0.47$ and $0.61$. \cite{2012JCAP...05..030S} found an $[25,50,75]$ percentiles of $[0.46, 0.52, 0.58]$, which is narrower than \cite{2015MNRAS.449.3171B}'s result, but matches their median of 0.54 quite well. This discrepancy will be discussed further in sections \ref{sec:discussion} and \ref{sec:conclusions}.  For the $a_2$ distribution, we found a median of 0.69 as opposed to the median of $0.67$ from the N-body simulations of \cite{2012JCAP...05..030S}. Our 25 and 75 percentiles are $0.57$ and $0.79$ respectively, while \cite{2012JCAP...05..030S} found values of $0.58$ and $0.78$. Therefore, this work overestimates the median value slightly and over predicts low $a_2$ values, but otherwise agrees quite well with simulation. It should also be noted that the distribution of $a_2$ was measured from a different simulation to which our model has not been tuned. Given that there are significant differences between axis ratios measured from various simulations (as will be discussed in \S\ref{sec:conditioned}), this agreement is evidence for the robustness of our model.

Finally, figure \ref{fig:triaxiality} shows the distribution of the triaxiality,
\begin{equation}
T=\frac{1 - a_2^2}{1 - a_3^2},
\end{equation}
of this model and compares it with the same parameter measured from N-body simulations by \cite{2012JCAP...05..030S}. The overall shape of the predicted distribution is similar to that found by \cite{2012JCAP...05..030S}, although the peak is offset to slightly smaller values of $T$.

\subsection{Conditioned Distributions and Mass Dependence\label{sec:conditioned}}

\begin{figure}
\includegraphics[width=85mm]{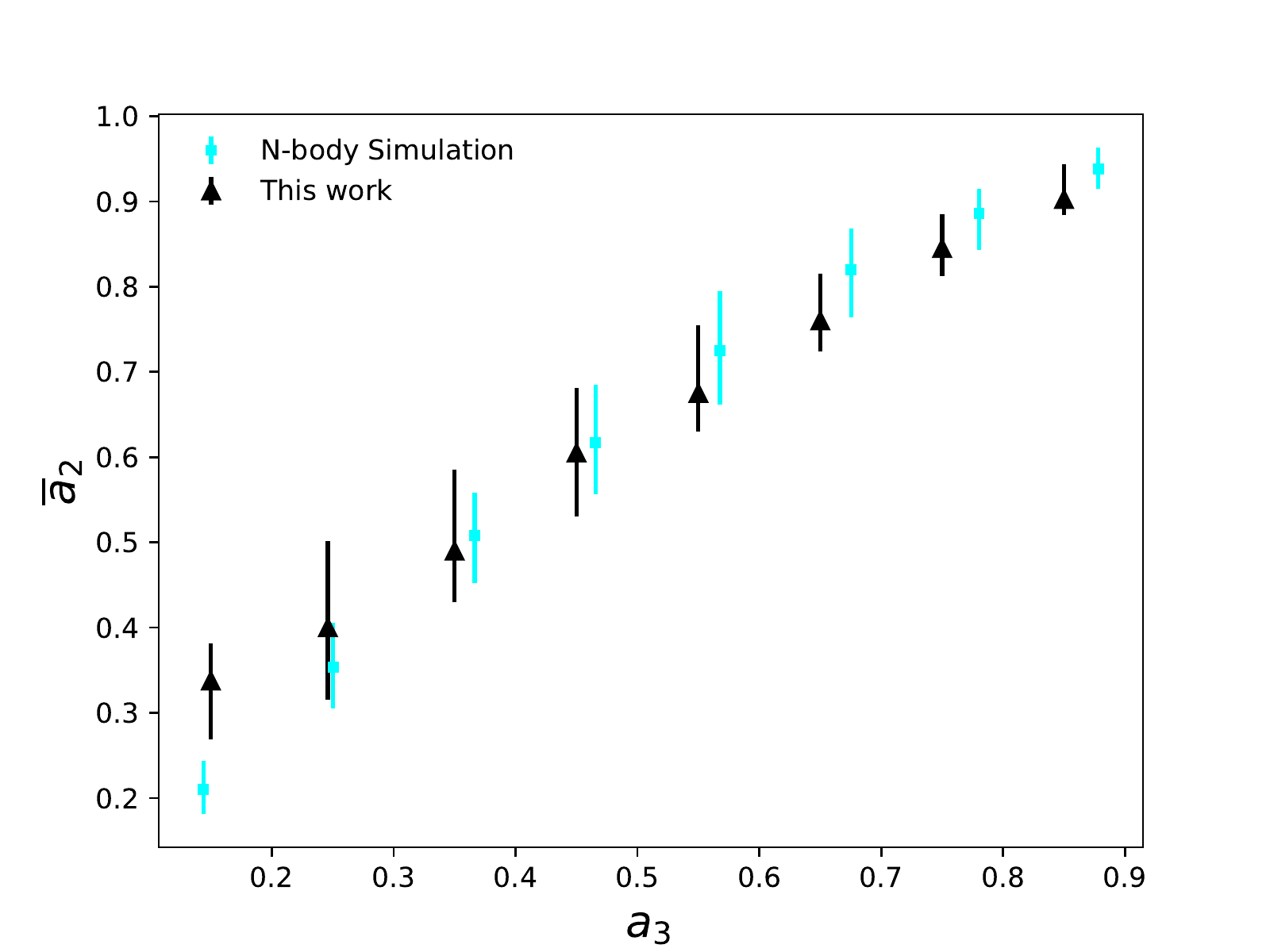}
\caption{The median value of $a_2$, $\bar{a}_2$, conditioned on $a_3$ for halos of mass $10^{14}\mathrm{M}_\odot$ at $z=0$. Blue points show results from this work found by subdividing our sample of 10,000 merger trees into bins of $a_3$. Black points show the medians reported by \protect\cite[][their Figure~12]{2015MNRAS.449.3171B} with error bars indicating the inter-quartile range in both data sets. \protect\cite{2015MNRAS.449.3171B} show results for a variety of mass intervals. Here we show their result for halo masses between $10^{13.4}$ and $10^{14.2} \mathrm{M}_\odot$.}
\label{fig:conditioned}
\end{figure}

In Figure~\ref{fig:conditioned} we examine the correlation between axis ratios predicted by our model by considering the median value of $a_2$ conditioned on the value of $a_3$, $\bar{a}_2(a_3)$. This is shown by the blue points, which were computed by subdividing our sample of 10,000 merger trees into bins of $a_3$, and then finding the median $a_2$ in each bin. For this calculation we again used trees with $z=0$ halo mass of $10^{14}\mathrm{M}_\odot$. For comparison we show N-body results from \cite{2015MNRAS.449.3171B} as the black points. \cite{2015MNRAS.449.3171B} show that the relation is largely independent of halo mass---nevertheless, we plot their result for halos in the mass range $10^{13.4}$ to $10^{14.2} \mathrm{M}_\odot$ to most closely match our halo masses. Our model matches the N-body results of \cite{2015MNRAS.449.3171B} very well for $a_3 > 0.3$. At lower value it somewhat overestimates the value of $\bar{a}_2(a_3)$.

\begin{figure}
\includegraphics[width=85mm]{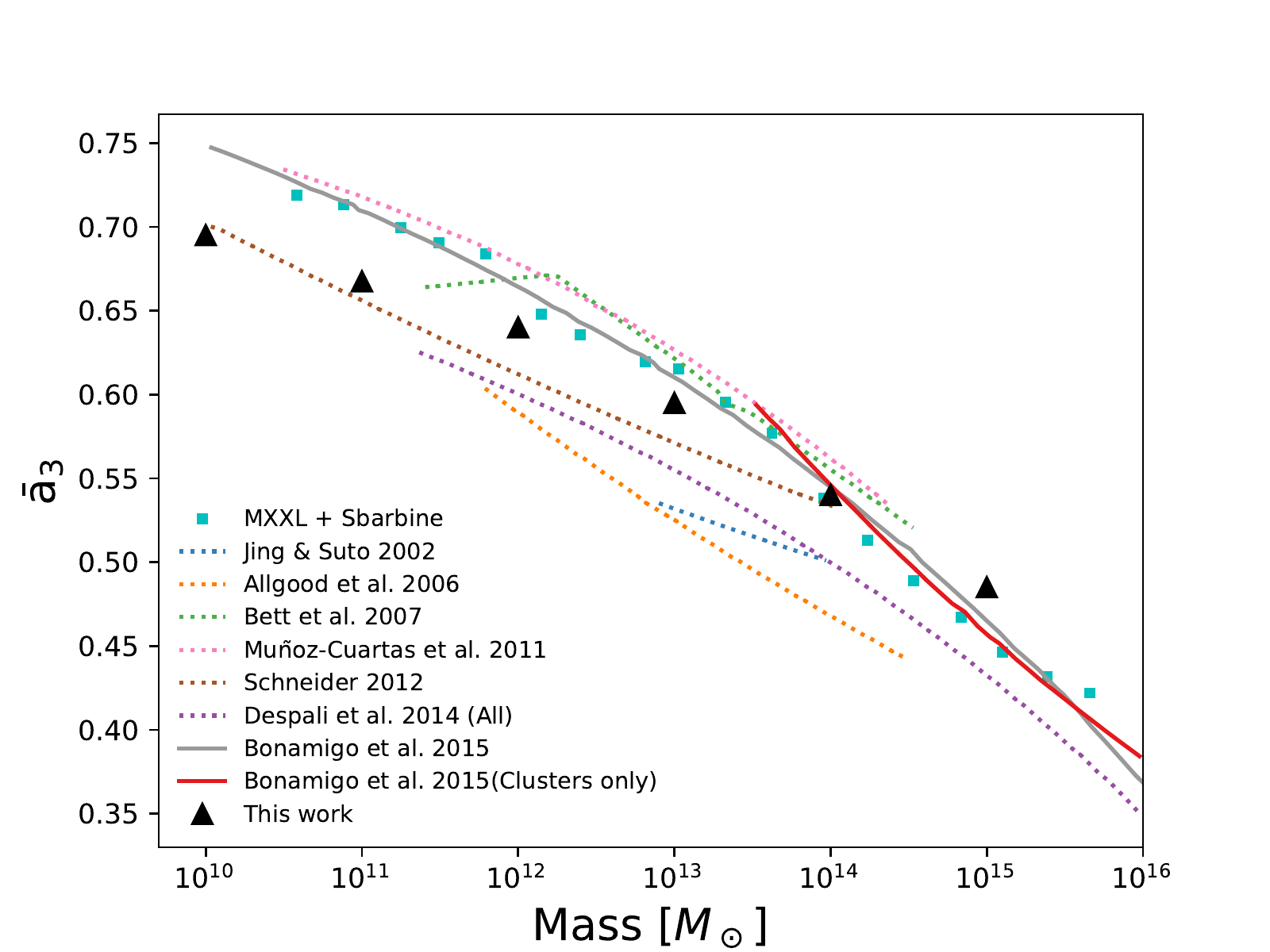}
\caption{The median value of $a_3$, $\bar{a}_3$, as a function of $z=0$ halo mass. Black triangles show results from this work found by applying our model to 10,000 merger tree realizations in each of six different halo mass bins. The various N-body curves are taken from \protect\citeauthor{2015MNRAS.449.3171B}~(\protect\citeyear{2015MNRAS.449.3171B}; Figure~13).}
\label{fig:massDependence}
\end{figure}

Finally, in Figure \ref{fig:massDependence}, we plot the median $a_3$, $\bar{a}_3$, as a function of $z=0$ halo mass. Black triangles show results from this work found by applying our model to 10,000 merger tree realizations in each of six different halo mass bins. Cyan squares show results from \cite{2015MNRAS.449.3171B} while lines show a compilation of other N-body results \citep{2015MNRAS.449.3171B,2006MNRAS.367.1781A,2007MNRAS.376..215B,2011MNRAS.411..584M,2012JCAP...05..030S,2014MNRAS.443.3208D,2002ApJ...574..538J} gathered by \cite{2015MNRAS.449.3171B}. Statistical uncertainties on the results from this work are at the sub-percent level---small compared the the difference between N-body results seen here. While the N-body results generally have small statistical uncertainty\footnote{\protect\cite{2015MNRAS.449.3171B} do not show error bars on their points, but the bin-to-bin variations suggest statistical variations of no more than $\pm 0.02$ in $\bar{a}_3$.}, there is a significant variation between the various N-body-derived results\footnote{This variation is likely due, at least in part, to the poor convergence properties of halo shapes in N-body simulations as shown, for example, by \protect\citeauthor{2021MNRAS.500.3309M}~(\citeyear{2021MNRAS.500.3309M}; Figure 3)---with significant differences between different simulations persisting even in halos resolved with $10^5$ particles.}. Nevertheless, all show the same general trend of decreasing $\bar{a}_3$ with increasing mass. Our model predicts a similar trend and lies within the spread of the N-body results. 

\section{Discussion}\label{sec:discussion}

Our model for the triaxial shape of dark matter halos functions by tracking the energy tensor of a halo during each merger event in its formation history, under the assumption that the energy tensor is approximately conserved. Minor corrections, which break perfect conservation of the energy tensor, are applied during each merger, and a slow sphericalization is allowed to occur between each merger event. We verify our model by comparing its expectations to idealized N-body simulations of merging, triaxial halo pairs, finding that in the majority of cases it matches the measured axis ratios with a median absolute deviation of $\Delta \approx 0.04$.

Applying this model to a merger tree, we are able to track the energy tensor and axis length ratios along the main branch of the tree\footnote{We in fact track these along all branches, although the results may be affected by resolution in lower mass branches.}, and record how the triaxial shape of halos changes over time.

We find that statistics of halo triaxial shape are very sensitive to the sphericalization parameter, $\alpha$, which controls the timescale for sphericalization. A value of $\alpha \sim 1$ is found to result in a good match to the properties of triaxial halos found in cosmological N-body simulations, suggesting that the orbital precession frequency (equation~\ref{eq:precession}) is the appropriate timescale for sphericalization. However, some of our idealized N-body experiments show more complex sphericalization behaviour. This is an area where improved understanding of the relaxation of triaxial halos could significantly improve our model.

Figure~\ref{fig:pdfUnconditioned} demonstrates that our model produces reasonable agreement with the distribution of axis length ratios measured in cosmological N-body simulation. The medians of the distributions from our model and the N-body simulation match quite well (once we have tuned the sphericalization parameter, $\alpha$), and the general shape is similar.  However, as one can see from the $a_2$ and $a_3$ histograms (Figure~\ref{fig:pdfUnconditioned}), our model produces a significantly more broad distribution than that from the N-body simulation. This is particularly apparent for very low $a_3$ values. As can be seen in Figure~\ref{fig:singleHistory}, low $a_2$ and $a_3$ values occur after a major merger. Therefore, it is possible that our model does not fully capture the dynamics of the largest mergers. For example, the $(1+M_2/M_1)^{-\gamma}$ term in equation~(\ref{eq:energyFinal}), introduced to account for the possibility of mass/energy ejection from the system during the merger, may be insufficient to capture the behaviour of the most massive mergers. Similarly, it is possible that the underlying assumption that the energy tensor is approximately conserved may simply fail for cosmological major mergers (although it holds well for mergers of idealized halos as discussed in \S\ref{sec:sphericalization}). A second possible cause for the excess of low $a_2$ and $a_3$ values is that our simple criterion for excluding unvirialized halos (based on the time since the last major merger) is insufficient to match in detail the relaxation criteria applied by \cite{2015MNRAS.449.3171B} to their sample. Other possible causes for this discrepancy include the limitations of the simple ``$(b,\gamma)$'' model\footnote{We have used this model unchanged from the form used by \protect\cite{johnson_random_2021} who applied this to a similar model for halo concentrations based on the scalar energy of halos. Since the trace of the energy tensor is the scalar energy, we expect that applying this same model here should ensure that we similarly match the behaviour of the trace of the energy tensor, but there is no guarantee that it will give the correct behaviour in a more general sense.} for non-conservation of the energy tensor in mergers, and our simple model for sphericalization\footnote{As already discussed, it is clear that the precession frequency given by equation~(\protect\ref{eq:precession}) is not the correct timescale for sphericalization. It is also possible that equation~(\protect\ref{eq:sphericalization}) is not the correct functional form for the evolution of the eigenvalues of the energy tensor during sphericalization.}, given by equation~(\ref{eq:sphericalization}).  As can be seen in Figure~\ref{fig:energyTensorSimulation}, the dynamics of mixing are quite complex, and sphericalization does not necessarily turn on immediately in the same way as Figure~\ref{fig:singleHistory}.

These limitations do not invalidate the underlying intent of the model, namely to describe the qualitative evolution of triaxial halo shapes measured in N-body simulations, and to provide an approximate quantitative description of them. They do, however, suggest a rich variety of improvements one could explore in future works.

To explore the predictive power of our model, once optimized to match unconditioned distributions of axis length ratios, we have considered the triaxiality parameter (Figure~\ref{fig:triaxiality}), conditioned distributions (Figure~\ref{fig:conditioned}), and mass-dependent distributions (Figure~\ref{fig:massDependence}). The distribution of triaxiality parameters is of the same shape as that found in N-body simulations, although offset to somewhat lower values, while we find that our model's predictions for conditioned distributions and mass dependence to both agree extremely well with measurements from N-body simulations. The median $a_2$ conditioned on $a_3$ shown in Figure~\ref{fig:conditioned} suggests that our model correctly captures the correlated evolution of $a_2$, $a_3$ during mergers---except perhaps at $a_3 < 0.3$ where we slightly overestimate $\bar{a}_2(a_3)$. Given the heavy tail of low $a_3$ values in \ref{fig:pdfUnconditioned}, this is perhaps due to the overpopulation of low $a_3$ halos in our model. Mass-dependence, which can only arise from the differences encoded into the structure of the merger trees, suggests that our model is correctly capturing the physics that drives triaxial evolution during halo assembly.  Improvements such as those discussed above could likely improve the  agreement at $M = 10^{15} \mathrm{M}_\odot$, but given the scatter between different N-body determinations at lower mass scales, we are hesitant to overfit to a single data set.

\section{Conclusion}\label{sec:conclusions}

The triaxial shapes of dark matter halos are important due to their consequences for the orbital evolution of subhalos, the strength of gravitational lensing produced by a halo, and many other factors.  This work builds on two previous papers \citep{benson_random-walk_2020,johnson_random_2021} which track halo spin and concentration using a similar approach. In this paper we expand their random-walk methodology to gain insight into the physical mechanisms governing the evolution of triaxial halo shapes. We find that the evolution of these triaxial halo shapes is driven by a combination of mergers and sphericalization. After calibration of our model to match the median $a_3$ in cluster halos, we find that these features are sufficient to explain conditioned distributions of axis ratios and mass dependence measured in cosmological N-body simulations, although the model underpredicts the width of the distributions of $a_2$ and $a_3$. While our model does not produce high-precision matches the N-body measurements, this was not the goal.  Instead, we aimed to explore whether thinking about triaxial halo shape evolution in this way is a productive avenue of exploration.
Future improvements, if desired, should focus on generalizing the correction factors (for environmental interactions, and sphericalization) employed here. 

There are a number of other avenues that could be explored in future work. Firstly, this work connects triaxial shape evolution directly to the growth and merger history of a halo. Connecting this back to the previous work with this framework would allow for predictions for the joint distributions of halo spin, concentration, and triaxial shape and how these grow together.  Second, our model only considers the global triaxial shape of halos, while radial variations are known to exist.  For example \cite{2011MNRAS.416.1377V} found that each radial shell of a halo retains a memory of its shape at the time at which it was assembled. Our model applies to halos, but could be adapted for subhalos. For these structures there are additional drivers of triaxial shape evolution, notably tidal forces from their host halo, which can cause subhalos to become more prolate \citep{2007ApJ...671.1135K}, but likely also lead to some additional sphericalization due to tidal heating effects. Finally, it may be interesting to extend our model, currently developed for cold dark matter, to cases where dark matter has self-interactions which is known to affect halo triaxiality \citep{2013MNRAS.430..105P,2018PhR...730....1T}. Such self-interactions could lead to more rapid thermalization and sphericalization which could be incorporated into our sphericalization model.

\section*{Acknowledgements}

We thank Xiaolong Du, Fangzhou Jiang, Ethan Nadler, and Shengqi Yang for helpful discussions and comments on this work. The CosmoSim database used in this paper is a service by the Leibniz-Institute for Astrophysics Potsdam (AIP). The MultiDark database was developed in cooperation with the Spanish MultiDark Consolider Project CSD2009-00064. The authors gratefully acknowledge the Gauss Centre for Supercomputing e.V. (www.gauss-centre.eu) and the Partnership for Advanced Supercomputing in Europe (PRACE, {\tt www.prace-ri.eu}) for funding the MultiDark simulation project by providing computing time on the GCS Supercomputer SuperMUC at Leibniz Supercomputing Centre (LRZ, {\tt www.lrz.de}).

\section*{Data Availability}

The data underlying this article will be shared on reasonable request to the corresponding author.


\bibliographystyle{mnras}
\bibliography{sample}

\appendix

\section{Test of Non-conservation}

\begin{figure}
\includegraphics[width=85mm]{images/a3_all_parameters_off.pdf}
\caption{Probability distribution functions for the axis ratio $a_3$ assuming a conserved energy tensor. Results were derived by applying our model to 10,000 realizations of merger trees with mass $10^{14}\mathrm{M}_\odot$ at $z=0$, and are compared to N-body simulation results from  \protect\cite{2015MNRAS.449.3171B}.}
\label{fig:parametersOff}
\end{figure}

A key assumption of this work is that the halo's energy tensor is not precisely conserved during its evolution. There is strong justification for this from N-body simulations, as discussed in \S\ref{sec:simulations}. However, it is interesting to explore what would happen to our model if these phenomena were ignored. As discussed in \S\ref{sec:beyondConservation}, non-conservation involves 3 tunable parameters, $(b,\gamma, \alpha).$  Figure \ref{fig:parametersOff} explores how essential these features are to the model by setting $b = \gamma = \alpha = 0.$  As expected, the resulting probability distribution function of $a_3$ now strongly disagrees with that measured from N-body simulations. Without these non-conservation effects, especially sphericalization, the axis ratios are biased too small, resulting in halos that are far too elongated. Clearly these corrections are a necessary component of our model---Figure \ref{fig:parametersOff} reinforces the result of \S\ref{sec:simulations} that the underlying phenomena of halo mergers are being successfully captured in this model.

\bsp	
\label{lastpage}
\end{document}